\documentclass[acmsmall,screen,balance]{acmart}
\usepackage{multirow}
\usepackage{xcolor}
\usepackage{adjustbox,pdflscape} 
\usepackage{cellspace} 
\usepackage{tabularray} 
\usepackage{float}
\usepackage[edges]{forest} 
\usepackage{url}
\usetikzlibrary{trees, positioning, shapes, shadows, arrows.meta}
\usepackage{anyfontsize} 
\usepackage{tikz}
\usepackage{svg}
\usepackage{longtable}
\usepackage{color,soul}
\usepackage[capitalise,nameinlink]{cleveref}
\usepackage{tabularx,enumitem}
\usepackage{subfig}

\usepackage[textsize=scriptsize,colorinlistoftodos]{todonotes}
\let\todonote\todo
\renewcommand{\todo}[2]{\todonote[inline,color=red!20]{TODO (#1): #2}}

\tikzset{parent/.style={align=center,text width=2cm,fill=green!20,rounded corners=2pt},
	child/.style={align=center,text width=2.8cm,fill=green!50,rounded corners=6pt},
	grandchild/.style={fill=pink!50,text width=2.3cm}
}

\makeatletter
\renewcommand\tabularxcolumn[1]{>{\@minipagetrue}p{#1}}
\makeatother

\setlist[itemize]{align=parleft,left=0pt..1em}

\begin{document}
	
	\title{Microservices-based Software Systems Reengineering: \\State-of-the-Art and Future Directions}

        \author{Thakshila Dilruksh}
        \email{timiyamohott@student.unimelb.edu.au}
        \orcid{0009-0002-7422-2303}
        \affiliation{%
          \institution{The University of Melbourne}
          \state{Victoria}
          \country{Australia}
          \postcode{3010}
        }

        \author{Artem Polyvyanyy}
        \affiliation{%
          \institution{The University of Melbourne}
          \country{Australia}
          \postcode{3010}
          }
        \email{artem.polyvyanyy@unimelb.edu.au}
        \orcid{0000-0002-7672-1643}

        \author{Rajkumar Buyya}
        \affiliation{%
          \institution{The University of Melbourne}
          \country{Australia}
          \postcode{3010}
          }
        \email{rbuyya@unimelb.edu.au}
        \orcid{0000-0001-9754-6496}

        \author{Alistair Barros}
        \affiliation{%
          \institution{Queensland University of Technology}
          \country{Australia}}
        \email{alistair.barros@qut.edu.au}
        \orcid{0000-0001-8980-6841}

        \author{Colin Fidge}
        \affiliation{%
          \institution{Queensland University of Technology}
          \country{Australia}}
        \email{c.fidge@qut.edu.au}
        \orcid{0000-0002-9410-7217}

	\begin{abstract}
		\sloppypar
        Designing software compatible with cloud-based Microservice Architectures (MSAs) is vital due to the performance, scalability, and availability limitations. 
        As the complexity of a system increases, it is subject to deprecation, difficulties in making updates, and risks in introducing defects when making changes. 
        Microservices are small, loosely coupled, highly cohesive units that interact to provide system functionalities. 
        We provide a comprehensive survey of current research into ways of identifying services in systems that can be redeployed as microservices. 
        Static, dynamic, and hybrid approaches have been explored. 
        While code analysis techniques dominate the area, dynamic and hybrid approaches remain open research topics.
	\end{abstract}
	
  \maketitle

\section{Introduction}

Modernizing software systems is essential to obtain the benefits of the latest technical capabilities~\cite{BennettLegacySystems}. 
Monolithic, legacy mainframe-based software systems are an increasingly obsolete technology that suffers from scalability, maintainability, availability, and efficiency problems~\cite{TaibiLPProcessMotivationsandIssues,FritzschMSMigrationInIndustry,NewmanBuildingMicroservices,fowlerLewisMicroservices}. 
Therefore, there is an imperative to modernize such systems to obtain better performance and improve the overall user experience~\cite{ArcelliCPPerformanceDrivenSoftwareRefactoring}.

Consequently, a wave of migration of monolithic software to object-oriented platforms was observed at the end of the previous millennium~\cite{LuciaOOPMigration}. Later, service-oriented architectures (SOAs) emerged, and legacy software systems began moving toward service-oriented architectures. In an SOA, software systems are modular, with distributed modules having clearly defined interfaces~\cite{serviceOrientedArchitecture}. As opposed to the logically related operations in an SOA, the microservice architectural style emerged promising to distribute applications via fine-grained, loosely coupled, and highly cohesive autonomous components communicating via well-defined, lightweight protocols managing local, synchronized databases, achieving high scalability, availability, and efficiency~\cite{NewmanBuildingMicroservices,fowlerLewisMicroservices}.

Microservices were first discussed in 2011~\cite{fowlerLewisMicroservices}. 
In addition to addressing the aforementioned drawbacks of conventional software architectures, microservices enable independent development and deployment of services, flexibility in horizontal scaling in the cloud environment, and support for efficient development team management~\cite{MazlamiExtractionOfMicroservices}. 
Due to their multiple advantages, companies like Google, Netflix, Amazon, Uber, and eBay upgraded to microservice-based systems.

Companies often have a substantial investment in their corporate business systems and cannot afford to simply redevelop them entirely. Instead, a legacy system can be converted into a microservice system by incrementally extracting microservices from it. This approach has several advantages. 
Firstly, it makes best use of the company's existing investment in the original system, which is often considerable and spans several decades. Secondly, the complexity of a legacy system and the effort, time to market, and resource constraints (e.g., human resources) required to reimplement it from scratch can be prohibitive. Finally, only certain system parts may be suitable for migration, while others cannot benefit from or even will degrade when moved to the new architectural style. For example, functionality that is infrequently used, such as annual financial reporting, is probably best implemented in the head office's mainframe. Hence, the ability to extract specific services for reengineering and redeployment as microservices while leaving other functionalities unchanged is essential.

This article is the first comprehensive review covering various aspects of redesigning monolithic software systems by extracting discrete functions from them that could be re-implemented as microservices, including service discovery approaches and techniques, tools that support reengineering, data used to inform migration processes, evaluation methods for the resulting microservice systems, and challenges and limitations of the existing reengineering approaches; refer to \cref{subsubsec:scope}. Thus, this work contributes to a better understanding of the concept of a microservice discovery in terms of software architectural properties and recommends future research directions in the area of migrating monolithic software systems to microservice architectures. 

The remainder of the paper proceeds as follows.
\Cref{sec:research:methodology} describes the research methodology followed in this work.
\Cref{sec:results} presents the results of this literature review. 
Finally, \cref{sec:discussion} and \cref{sec:conclusion} discuss the results and state concluding remarks, respectively.

\section{Systematic Literature Review Process}
\label{sec:research:methodology}

In this work, we followed the guidelines for performing systematic literature review in software engineering proposed by \citet{KitchenhamC07} and further refined by \citet{KitchenhamB13}. 
Specifically, the review was conducted in three iterative phases: planning, executing, and reporting the review.
The planning and execution phases are reported in \cref{subsec:planning:the:review,subsec:executing:the:review}, respectively, while the review results are reported in \cref{sec:results,sec:discussion}.

\subsection{Planning the Review}
\label{subsec:planning:the:review}

The planning phase of the review process defines the research questions addressed in this work and develops the review protocol that specifies concrete steps for searching and selecting relevant primary studies in this literature review.
Next, we justify the scope of this literature review, see \cref{subsubsec:scope}.
Then, in \cref{subsubsec:research:questions}, we present research questions that cover the identified scope.
Finally, \cref{subsubsec:protocol} discusses the review protocol.

\subsubsection{Scope}
\label{subsubsec:scope}

Due to the many advantages of microservice-based architectures, reengineering monolithic software systems into microservices has gained attention.
Several early studies have been conducted to systematize existing works in the area~\cite{Schmidt,Schroer,Cojocaru,Kazanavicius,FritzschMSMigrationInIndustry,Capuano,Ponce,Jonas,Taibi,Carvalho,Wolfart,Manel}.
These studies were identified by first performing an initial search for survey and literature review papers in the area of interest and then including all additional secondary studies identified when searching for the relevant primary studies.
They explore different system modeling and microservice extraction techniques and criteria for evaluating the resulting microservices.

The existing literature reviews on migrating software systems into microservices are listed below.

\smallskip
\newcounter{litreview}
\renewcommand{\thelitreview}{LR\arabic{litreview}}
\begin{list}{LR\arabic{litreview}}{\usecounter{litreview}
\setlength{\itemsep}{0pt}
\setlength{\parsep}{0pt}
\setlength{\labelwidth}{3em}}
\item\label{lrSchmidt}
\citet{Schmidt} reviewed model-driven engineering and artifact-driven analysis approaches to identify potential microservices.
\item\label{lrSchroer}
\citet{Schroer} focused their survey on the techniques for identifying microservices during the software development process and evaluating identified microservices.
\item\label{lrCojocaru}
\citet{Cojocaru} discussed existing quality assessment criteria for microservices produced by semi-automatic migration tools and techniques.
\item\label{lrKazanavicius}
\citet{Kazanavicius} discussed microservice architectures and infrastructure requirements for microservices and compared six legacy system migration projects.
\item\label{lrFritzsch}
An analysis of migration processes from monolithic architectures to microservices for real-world industry systems was performed by \citet{FritzschMSMigrationInIndustry}, including the intentions, migration strategies, and challenges encountered by the companies during migration.
\item\label{lrCapuano}
Quality-driven approaches in migration, quality attributes analysis, and quality-driven process implementation were reviewed by \citet{Capuano}.
\item\label{lrPonce}
\citet{Ponce} conducted a literature study of migration techniques, the types of systems to which the proposed techniques are applied, methods for validating the migration techniques, and the challenges associated with such migrations.
\item\label{lrJonas}
\citet{Jonas} studied existing architectural refactoring approaches in the context of decomposing a monolithic application architecture into microservices and how they can be classified concerning the techniques and strategies used.
\item\label{lrTaibi}
A systematic mapping study on microservice-based architectural patterns and the advantages and disadvantages of those patterns was conducted by \Citet{Taibi}.
\item\label{lrCarvalho}
An exploratory online survey with 15 specialists experienced in migrating systems to microservices was accomplished by \citet{Carvalho} to analyze the criteria adopted by the industry for extracting microservices.
\item\label{lrWolfart}
The approaches to modernizing legacy software were discussed by \citet{Wolfart}.
\item\label{lrManel}
A taxonomy of service identification approaches that combines the inputs used for service identification, the process followed, the output of service identification, and the usability of service identification was developed by \citet{Manel}.
\end{list}

The bibliographic information on all the listed studies is provided in \cref{appendix:litReviewStudies}.

\Cref{table:studyComparison} compares the existing literature reviews. 
If a study has declared a specific review period or year, it is specified in the review period/year row.
Otherwise, the study year has been provided to indicate that the review period cannot go beyond that year.
If a study mentions the number of reviewed papers, it is indicated in the number of reviewed papers row.

\newcommand*\emptycirc[1][1.5ex]{\trimbox{0cm 0.1cm 0cm -0.1cm}{\tikz\draw (0,0) circle (#1);}} 
\newcommand*\halfcirc[1][1.5ex]{\trimbox{0cm 0.1cm 0cm -0.1cm}{%
  \begin{tikzpicture}
  \draw[fill] (0,0)-- (90:#1) arc (90:270:#1) -- cycle ;
  \draw (0,0) circle (#1);
  \end{tikzpicture}}}
\newcommand*\fullcirc[1][1.5ex]{\trimbox{0cm 0.1cm 0cm -0.1cm}{\tikz\fill (0,0) circle (#1);}} 

\begin{table}[t]
\caption{A comparison of existing literature reviews on reengineering of software systems into microservices}
\vspace{-3mm}
\adjustbox{max width=\textwidth}{%
\begin{tabular}{lcccccccccccccc|}
\hline
\multicolumn{1}{|l|}{\textbf{Literature review}} & \multicolumn{1}{c|}{\ref{lrSchmidt}} & \multicolumn{1}{c|}{\ref{lrSchroer}} & \multicolumn{1}{c|}{\ref{lrCojocaru}} & \multicolumn{1}{c|}{\ref{lrKazanavicius}} & \multicolumn{1}{c|}{\ref{lrFritzsch}} & \multicolumn{1}{c|}{\ref{lrCapuano}} & \multicolumn{1}{c|}{\ref{lrPonce}} & \multicolumn{1}{c|}{\ref{lrJonas}} & \multicolumn{1}{c|}{\ref{lrTaibi}} & \multicolumn{1}{c|}{\ref{lrCarvalho}} & \multicolumn{1}{c|}{\ref{lrWolfart}} & \multicolumn{1}{c|}{\ref{lrManel}} & \multicolumn{1}{c|}{\begin{tabular}[c]{@{}c@{}}This\\study\end{tabular}} & \multirow{3}{*}{\begin{tabular}[c]{@{}c@{}}\textbf{Research}\\\textbf{question(s)}\\\textbf{addressed}\\\textbf{in this}\\\textbf{study}\end{tabular}} \\ 
\cline{1-14}
\multicolumn{1}{|l|}{\textbf{Review period/year}} & \multicolumn{1}{c|}{\begin{tabular}[c]{@{}c@{}}2013--\\ 2019\end{tabular}} & \multicolumn{1}{c|}{2020} & \multicolumn{1}{c|}{\begin{tabular}[c]{@{}l@{}}1998--\\2018\end{tabular}} & \multicolumn{1}{c|}{2019} & \multicolumn{1}{c|}{2019} & \multicolumn{1}{c|}{\begin{tabular}[c]{@{}c@{}}2016--\\2022\end{tabular}} & \multicolumn{1}{c|}{2019} & \multicolumn{1}{c|}{2018} & \multicolumn{1}{c|}{2018} & \multicolumn{1}{c|}{2019} & \multicolumn{1}{c|}{2021} & \multicolumn{1}{c|}{2020} & \multicolumn{1}{c|}{\begin{tabular}[c]{@{}c@{}}2023\end{tabular}} & \\ 
\cline{1-14}
\multicolumn{1}{|l|}{\textbf{Number of reviewed papers}\begin{tabular}[c]{@{}c@{}}\,\\ \,\end{tabular}} & \multicolumn{1}{c|}{27} & \multicolumn{1}{c|}{31} & \multicolumn{1}{c|}{29} & \multicolumn{1}{c|}{N/A} & \multicolumn{1}{c|}{14} & \multicolumn{1}{c|}{58} & \multicolumn{1}{c|}{20} & \multicolumn{1}{c|}{10} & \multicolumn{1}{c|}{41} & \multicolumn{1}{c|}{15} & \multicolumn{1}{c|}{62} & \multicolumn{1}{c|}{41} & \multicolumn{1}{c|}{92} & \\ 
\hline\hline
\multicolumn{15}{|c|}{\textbf{Comparison of research questions}} \\ 
\hline\hline
\multicolumn{1}{|l|}{\begin{tabular}[c]{@{}l@{}}What are the techniques/approaches/patterns for legacy\\software reengineering?\end{tabular}} & \multicolumn{1}{c|}{\halfcirc} &\multicolumn{1}{c|}{\fullcirc} & \multicolumn{1}{c|}{\emptycirc} & \multicolumn{1}{c|}{\fullcirc} & \multicolumn{1}{c|}{\fullcirc} & \multicolumn{1}{c|}{\emptycirc} & \multicolumn{1}{c|}{\fullcirc} & \multicolumn{1}{c|}{\fullcirc} & \multicolumn{1}{c|}{\fullcirc} & \multicolumn{1}{c|}{\emptycirc} & \multicolumn{1}{c|}{\fullcirc} & \multicolumn{1}{c|}{\fullcirc} & \multicolumn{1}{c|}{\fullcirc} & \multicolumn{1}{c|}{\ref{rqtwoone}\&\ref{rqtwothree}} \\ \hline
    
    \multicolumn{1}{|l|}{\begin{tabular}[c]{@{}l@{}}What types of systems have the existing reengineering\\techniques been applied to?\end{tabular}} & \multicolumn{1}{c|}{\emptycirc} & \multicolumn{1}{c|}{\emptycirc} & \multicolumn{1}{c|}{\emptycirc} & \multicolumn{1}{c|}{\emptycirc} & \multicolumn{1}{c|}{\emptycirc} & \multicolumn{1}{c|}{\emptycirc} & \multicolumn{1}{c|}{\fullcirc} & \multicolumn{1}{c|}{\fullcirc} & \multicolumn{1}{c|}{\emptycirc} & \multicolumn{1}{c|}{\emptycirc} & \multicolumn{1}{c|}{\emptycirc} & \multicolumn{1}{c|}{\emptycirc} & \multicolumn{1}{c|}{\fullcirc} & \multicolumn{1}{c|}{\ref{rqtwoone}} \\ \hline
    
    \multicolumn{1}{|l|}{\begin{tabular}[c]{@{}l@{}}What tools are used for reengineering monolithic systems\\into microservices?\end{tabular}} & \multicolumn{1}{c|}{\emptycirc} & \multicolumn{1}{c|}{\emptycirc} & \multicolumn{1}{c|}{\emptycirc} & \multicolumn{1}{c|}{\emptycirc} & \multicolumn{1}{c|}{\emptycirc} & \multicolumn{1}{c|}{\emptycirc} & \multicolumn{1}{c|}{\emptycirc} & \multicolumn{1}{c|}{\emptycirc} & \multicolumn{1}{c|}{\emptycirc} & \multicolumn{1}{c|}{\fullcirc} & \multicolumn{1}{c|}{\emptycirc} & \multicolumn{1}{c|}{\emptycirc} & \multicolumn{1}{c|}{\fullcirc} & \multicolumn{1}{c|}{\ref{rqtwotwo}} \\ \hline
    
    \multicolumn{1}{|l|}{\begin{tabular}[c]{@{}l@{}}What inputs/outputs are used by the existing\\reengineering techniques?\end{tabular}} & \multicolumn{1}{c|}{\emptycirc} & \multicolumn{1}{c|}{\emptycirc} & \multicolumn{1}{c|}{\emptycirc} & \multicolumn{1}{c|}{\emptycirc} & \multicolumn{1}{c|}{\emptycirc} & \multicolumn{1}{c|}{\emptycirc} & \multicolumn{1}{c|}{\emptycirc} & \multicolumn{1}{c|}{\emptycirc} & \multicolumn{1}{c|}{\emptycirc} & \multicolumn{1}{c|}{\emptycirc} & \multicolumn{1}{c|}{\fullcirc} & \multicolumn{1}{c|}{\fullcirc} & \multicolumn{1}{c|}{\fullcirc} & \multicolumn{1}{c|}{\ref{rqtwofour}} \\ \hline
    
    \multicolumn{1}{|l|}{\begin{tabular}[c]{@{}l@{}}What evaluation criteria are used for the identified\\microservices?\end{tabular}} & \multicolumn{1}{c|}{\emptycirc} & \multicolumn{1}{c|}{\fullcirc} & \multicolumn{1}{c|}{\emptycirc} & \multicolumn{1}{c|}{\emptycirc} & \multicolumn{1}{c|}{\emptycirc} & \multicolumn{1}{c|}{\emptycirc} & \multicolumn{1}{c|}{\emptycirc} & \multicolumn{1}{c|}{\emptycirc} & \multicolumn{1}{c|}{\emptycirc} & \multicolumn{1}{c|}{\emptycirc} & \multicolumn{1}{c|}{\emptycirc} & \multicolumn{1}{c|}{\emptycirc} & \multicolumn{1}{c|}{\fullcirc} & \multicolumn{1}{c|}{\ref{rqtwofive}} \\ \hline
    
    \multicolumn{1}{|l|}{\begin{tabular}[c]{@{}l@{}}How reengineering processes/techniques are\\validated?\end{tabular}} & \multicolumn{1}{c|}{\emptycirc} & \multicolumn{1}{c|}{\emptycirc} & \multicolumn{1}{c|}{\emptycirc} & \multicolumn{1}{c|}{\emptycirc} & \multicolumn{1}{c|}{\emptycirc} & \multicolumn{1}{c|}{\emptycirc} & \multicolumn{1}{c|}{\fullcirc} & \multicolumn{1}{c|}{\emptycirc} & \multicolumn{1}{c|}{\emptycirc} & \multicolumn{1}{c|}{\emptycirc} & \multicolumn{1}{c|}{\emptycirc} & \multicolumn{1}{c|}{\emptycirc} & \multicolumn{1}{c|}{\fullcirc} & \multicolumn{1}{c|}{\ref{rqtwofive}} \\ \hline
    
    \multicolumn{1}{|l|}{\begin{tabular}[c]{@{}l@{}}What quality-driven/assessment criteria are used for\\reengineering?\end{tabular}} & \multicolumn{1}{c|}{\emptycirc} & \multicolumn{1}{c|}{\emptycirc} & \multicolumn{1}{c|}{\fullcirc} & \multicolumn{1}{c|}{\emptycirc} & \multicolumn{1}{c|}{\emptycirc} & \multicolumn{1}{c|}{\fullcirc} & \multicolumn{1}{c|}{\emptycirc} & \multicolumn{1}{c|}{\emptycirc} & \multicolumn{1}{c|}{\emptycirc} & \multicolumn{1}{c|}{\emptycirc} & \multicolumn{1}{c|}{\emptycirc} & \multicolumn{1}{c|}{\emptycirc} & \multicolumn{1}{c|}{\fullcirc} & \multicolumn{1}{c|}{\ref{rqtwofive}} \\ \hline
    
    \multicolumn{1}{|l|}{\begin{tabular}[c]{@{}l@{}}What quality attributes are analyzed, and how have they\\been implemented for reengineering?\end{tabular}} & \multicolumn{1}{c|}{\emptycirc} & \multicolumn{1}{c|}{\emptycirc} & \multicolumn{1}{c|}{\fullcirc} & \multicolumn{1}{c|}{\emptycirc} & \multicolumn{1}{c|}{\emptycirc} & \multicolumn{1}{c|}{\fullcirc} & \multicolumn{1}{c|}{\emptycirc} & \multicolumn{1}{c|}{\emptycirc} & \multicolumn{1}{c|}{\emptycirc} & \multicolumn{1}{c|}{\emptycirc} & \multicolumn{1}{c|}{\emptycirc} & \multicolumn{1}{c|}{\emptycirc} & \multicolumn{1}{c|}{\fullcirc} & \multicolumn{1}{c|}{\ref{rqtwofive}} \\ \hline
    
    \multicolumn{1}{|l|}{\begin{tabular}[c]{@{}l@{}}What are the challenges of reengineering legacy\\software systems into microservices?\end{tabular}} & \multicolumn{1}{c|}{\emptycirc} & \multicolumn{1}{c|}{\emptycirc} & \multicolumn{1}{c|}{\emptycirc} & \multicolumn{1}{c|}{\fullcirc} & \multicolumn{1}{c|}{\fullcirc} & \multicolumn{1}{c|}{\emptycirc} & \multicolumn{1}{c|}{\fullcirc} & \multicolumn{1}{c|}{\emptycirc} & \multicolumn{1}{c|}{\emptycirc} & \multicolumn{1}{c|}{\emptycirc} & \multicolumn{1}{c|}{\emptycirc} & \multicolumn{1}{c|}{\emptycirc} & \multicolumn{1}{c|}{\fullcirc} & 
    \multicolumn{1}{c|}{\ref{rqthree}} \\ \hline
    
    \multicolumn{1}{|l|}{\begin{tabular}[c]{@{}l@{}}What usability aspects, advantages, and disadvantages/\\limitations are highlighted?\end{tabular}} & \multicolumn{1}{c|}{\emptycirc} & \multicolumn{1}{c|}{\emptycirc} & \multicolumn{1}{c|}{\emptycirc} & \multicolumn{1}{c|}{\emptycirc} & \multicolumn{1}{c|}{\emptycirc} & \multicolumn{1}{c|}{\emptycirc} & \multicolumn{1}{c|}{\emptycirc} & \multicolumn{1}{c|}{\emptycirc} & \multicolumn{1}{c|}{\fullcirc} & \multicolumn{1}{c|}{\fullcirc} & \multicolumn{1}{c|}{\emptycirc} & \multicolumn{1}{c|}{\fullcirc} & \multicolumn{1}{c|}{\fullcirc} & 
    \multicolumn{1}{c|}{\ref{rqthree}} \\ \hline
    
    \multicolumn{1}{|l|}{\begin{tabular}[c]{@{}l@{}}What are the intentions for reengineering software\\systems into microservices?\end{tabular}} & \multicolumn{1}{c|}{\emptycirc} & \multicolumn{1}{c|}{\emptycirc} & \multicolumn{1}{c|}{\emptycirc} & \multicolumn{1}{c|}{\emptycirc} & \multicolumn{1}{c|}{\fullcirc} & \multicolumn{1}{c|}{\emptycirc} & \multicolumn{1}{c|}{\emptycirc} & \multicolumn{1}{c|}{\emptycirc} & \multicolumn{1}{c|}{\emptycirc} & \multicolumn{1}{c|}{\emptycirc} & \multicolumn{1}{c|}{\emptycirc} & \multicolumn{1}{c|}{\emptycirc} & \multicolumn{1}{c|}{\emptycirc} & 
    \multicolumn{1}{c|}{N/A} \\ \hline
    
    \multicolumn{1}{|l|}{\begin{tabular}[c]{@{}l@{}}What are the roles and responsibilities involved in\\the identification of microservices?\end{tabular}} &  \multicolumn{1}{c|}{\fullcirc} & \multicolumn{1}{c|}{\emptycirc} & \multicolumn{1}{c|}{\emptycirc} & \multicolumn{1}{c|}{\emptycirc} & \multicolumn{1}{c|}{\emptycirc} & \multicolumn{1}{c|}{\emptycirc} & \multicolumn{1}{c|}{\emptycirc} & \multicolumn{1}{c|}{\emptycirc} & \multicolumn{1}{c|}{\emptycirc} & \multicolumn{1}{c|}{\emptycirc} & \multicolumn{1}{c|}{\emptycirc} & \multicolumn{1}{c|}{\emptycirc} & \multicolumn{1}{c|}{\emptycirc} &
    \multicolumn{1}{c|}{N/A} \\ \hline
\multicolumn{15}{c}{\vspace{20mm}\fullcirc\;{Addressed}\quad\quad\halfcirc\;{Partially addressed}\quad\quad\emptycirc\;{Not addressed}}
\end{tabular}
}
\label{table:studyComparison}
\vspace{-12mm}
\end{table}

The research questions listed in the first column of \cref{table:studyComparison}  are the research questions addressed in the existing studies.
We merged similar research questions and rephrased them to ensure the consistent use of terminology. 
The table summarizes which research questions are fully, partially, or not addressed in the existing literature reviews.
\ref{lrSchmidt} focuses on semi-automated approaches to reengineering and, thus, addresses the question of what techniques/approaches/patterns are used for legacy software reengineering only partially. \Cref{table:studyComparison} confirms that existing literature reviews are limited in scope, objectives, and coverage. 
It is, therefore, essential to analyze and systematize existing works comprehensively, cross-cutting different techniques, system modeling approaches, and evaluation strategies to understand the state-of-the-art, research gaps, and promising avenues for future work.
Hence, this work to address the gap.
	
\subsubsection{Research Questions}
\label{subsubsec:research:questions}

This literature review was conducted to examine existing methods, techniques, and tools for reengineering software systems into microservices, understand the limitations of the existing approaches, and identify fruitful avenues for future work. 
Consequently, we formulated the following research questions to guide our study.

\smallskip
\newcounter{researchquestion}
\renewcommand{\theresearchquestion}{RQ\arabic{researchquestion}}
\newcounter{researchquestiontwo}
\renewcommand{\theresearchquestiontwo}{RQ2.\arabic{researchquestiontwo}}
\begin{list}{RQ\arabic{researchquestion}}{\usecounter{researchquestion}
\setlength{\itemsep}{0pt}
\setlength{\parsep}{0pt}
\setlength{\labelwidth}{3em}}
\item\label{rqone}
How did research on the reengineering of software systems into micro\-service-based systems develop over time?
\item\label{rqtwo}
What approaches are used to reengineer software systems into micro\-service-based systems, and how are reengineered systems evaluated?
		\begin{list}{RQ2.\arabic{researchquestiontwo}}{\usecounter{researchquestiontwo}
		\setlength{\itemsep}{0pt}
		\setlength{\parsep}{0pt}
		\setlength{\labelwidth}{3em}
		\setlength{\itemindent}{4mm}}
		\item\label{rqtwoone}
		What classes of approaches (e.g., static and dynamic) exist?
		\item\label{rqtwotwo}
		What tools exist, and which level of automation do they support?
		\item\label{rqtwothree}
		Which techniques/algorithms are used?
		\item\label{rqtwofour}
		How is data (e.g., software logs) are used?
		\item\label{rqtwofive}
		How are the reengineered systems evaluated?
		\end{list}
\item\label{rqthree}
What are the challenges and limitations of existing methods for reengineering software systems into micro\-service-based systems?
\end{list}

\noindent
The research questions were defined to maximize the coverage of the questions addressed in the early studies (cf.\ the first column in \cref{table:studyComparison}) and to understand and refine them further.
The last column in \cref{table:studyComparison} maps the research questions addressed in this work onto the questions studied elsewhere. 
Our study does not consider the last two research questions listed in the table. 
Due to the well-confirmed benefits of reengineering legacy systems into microservices and the typical roles involved in the software development lifecycle, we excluded these two aspects from our study.


	
\subsubsection{Protocol}
\label{subsubsec:protocol}

All the publications analyzed in this study were retrieved from five databases widely used to index publications in the areas of computer science and software engineering: 
Web of Science,\footnote{\url{https://clarivate.com/webofsciencegroup/solutions/web-of-science}}
Scopus,\footnote{\url{https://www.scopus.com/}}
ScienceDirect,\footnote{\url{https://www.sciencedirect.com/}}
ACM Digital Library,\footnote{\url{http://portal.acm.org/}} and 
IEEE Xplorer Digital Library.\footnote{\url{https://ieeexplore.ieee.org/}}
These databases provide good coverage of primary sources from high-quality academic journals and peer-reviewed conferences~\cite{Gusenbauer}.

To maximize the chances of discovering papers that can contribute to answering the research questions defined in this study, we defined these keywords: ``microservice'', ``reengineer'', ``redesign'', ``discover'', and ``identify''.
The ``microservice'' keyword was defined as the study targets microservices systems.
As this work focuses on reengineering software systems into microservice-based systems, the keywords ``reengineer'' and ``redesign'' were included.
Finally, the keywords ``discover'' and ``identify'' were formulated since we are concerned about identifying microservices.

The types of searched manuscripts were limited to journal articles and conference proceedings papers written in English. 
The paper category was limited to computer science. 
For instance, the search query we used for the Web of Science database is given below.

\begin{quote}
(TS~= (microservice* AND (reengineer* OR re-engineer* OR redesign* OR re-design* OR discover* OR identify*))) AND  (WC~= (Computer Science)) AND (DT~= (Article OR Proceedings Paper)) AND (LA~= (English))
\end{quote}

\noindent
This query was translated into similar queries for the other four databases and these queries are listed in \cref{appendix:queries}.
To guide the identification of primary studies by supporting the decisions on which publication to review in our study, we defined inclusion and exclusion criteria listed in \cref{table:inclusionExclusion}.
These criteria were used to identify the suitability of the primary studies.

\begin{table}
\footnotesize
\caption{Inclusion and exclusion criteria}
\label{table:inclusionExclusion}
\vspace{-3mm}
\begin{tabularx}{\textwidth}[h]{l|X}
\toprule
Criterion type & Criterion definition\\ 
\midrule
Inclusion & 
\begin{enumerate}[wide = 0pt, leftmargin =*, noitemsep, topsep=0pt, before=\vspace*{-2mm}, after=\vspace*{-\dimexpr\baselineskip}]
\item Study is on legacy software system reengineering
\item Study is on requirements for reengineering of legacy software systems
\item Study is on a technique for evaluating functional consistency of a reengineered software system
\item Study is on a technique for evaluating the performance of a microservice system
\item Study is on using software logs for legacy software system reengineering
\item Study is on an approach for evaluating microservices
\end{enumerate}\\
\midrule
Exclusion & 
\begin{enumerate}[wide = 0pt, leftmargin =*, noitemsep, topsep=0pt, before=\vspace*{-2mm}, after=\vspace*{-\dimexpr\baselineskip}]
\item Study is not related to software systems
\item Study is on microservice system deployment, self-adjusting models, Quality of Service, and scalability
\item Study is on networks or load testing, security, and fault tolerance of software systems
\item Study does not present sufficient technical details to contribute to at least one research question addressed in this literature review
\item Study did not undergo a peer-review process, for example, published in a non-reviewed journal or conference papers, theses, books and book chapters, and doctoral dissertations
\item Study is a secondary literature review
\item Study is not in English
\end{enumerate}\\
\bottomrule
\end{tabularx}
\vspace{-3mm}
\end{table}
	
\subsection{Executing the Review}
\label{subsec:executing:the:review}

\begin{figure}[t]
\centering
\includegraphics[width=0.9\textwidth]{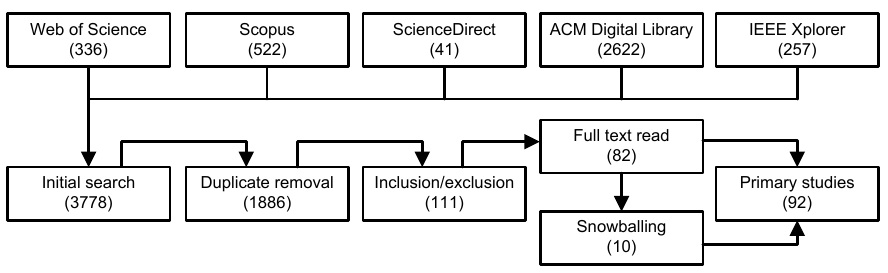}
\vspace{-4mm}
\caption{Overview of the search process}
\label{fig:searchProcess}
\vspace{-4mm}
\end{figure}

\Cref{fig:searchProcess} summarizes the search process for identifying primary studies followed in this work, including the accomplished search stages and the number of papers identified in these stages.
The initial search for relevant papers over the five databases was conducted on the 23\textsuperscript{rd} of January 2023. 
To ensure the full coverage of works relevant to this study on the date the initial search was conducted, we did not impose restrictions on the publication dates of the retrieved references. 
In this initial search, 3\,778 references were retrieved.
As a paper can be indexed by several databases, we removed duplicate references to result in 1\,886 distinct references.
To determine their relevance to our study, all the references were evaluated against the inclusion and exclusion criteria from \Cref{table:inclusionExclusion} using a checklist-based scoring procedure.
Papers on legacy system refactoring, requirements for refactoring, refactoring techniques, and evaluation of reengineered systems were included for further analysis. 
Studies not related to our research questions, for example, papers on networks and deployment of micro\-service-based systems, non-peer-reviewed studies, studies not related to software systems, or not in English, were excluded from further processing. 
At the end of this stage, 111 papers were identified as potentially relevant for our literature review.
The inclusion/exclusion decisions were taken based on paper titles and abstracts.
Hence, papers with unclear exclusion decisions were kept for further full text analysis.
The full text read of 111 papers revealed 82 relevant studies.
During the review of the papers selected for full-text analysis, relevant references were noted.
These references were analyzed in the snowballing stage, and relevant works were included in the study. 
Both forward and backward searching on references were performed. 
Ten additional papers were included in the snowballing stage.
Consequently, the presented search process has resulted in the identification of 92 primary studies.


The decisions that governed the selection of primary studies based on the initial search results were carried out independently by the first two authors of this paper and then consolidated into final decisions.
The identified 92 primary studies were analyzed in-depth, and relevant insights were extracted from them.
To systematize the knowledge extracted during the in-depth analysis of the primary studies, we followed a method for taxonomy development by \citet{Nickerson2013}.
It is an iterative approach to identifying concepts and their characteristics and grouping them into dimensions.
The method guides the evaluation of the developed taxonomies for usefulness, like the completeness of taxonomy dimensions and robustness. 

\section{Results of the Literature Review}
\label{sec:results}

This section elaborates on the findings of the literature review based on the research questions. \Cref{appendix:studyList} shows the list of studies related to legacy system migration. \Cref{table:selectedStudiesClassificationTable} summarizes the classification of papers from the selected 92 studies. The majority of the papers (63\%) explain legacy system migration strategies, whereas most of the remainder (32\%) focus on industry/architectural reviews, specifically case studies only. A small number (4\%) discussed greenfield development where new system implementation in a microservice-based architecture is considered.
	
\begin{table}[ht]
\captionof{table}{Paper classification details}
\centering
\begin{tabular}{|l |c|}
\hline
\textbf{Type of study} & \textbf{Number of papers} \\
\hline
Software system migration frameworks & 58 \\
\hline
Case studies and industry interviews & 30 \\
\hline
Greenfield development & 4 \\
\hline
\end{tabular}
\label{table:selectedStudiesClassificationTable}
\end{table}

\subsection{RQ1) How did research on the reengineering of software systems into microservice-based systems develop over time?}

\begin{figure}[ht]
\centering
\subfloat[\centering Level of automation over time ]{{\includegraphics[width=6.5cm]{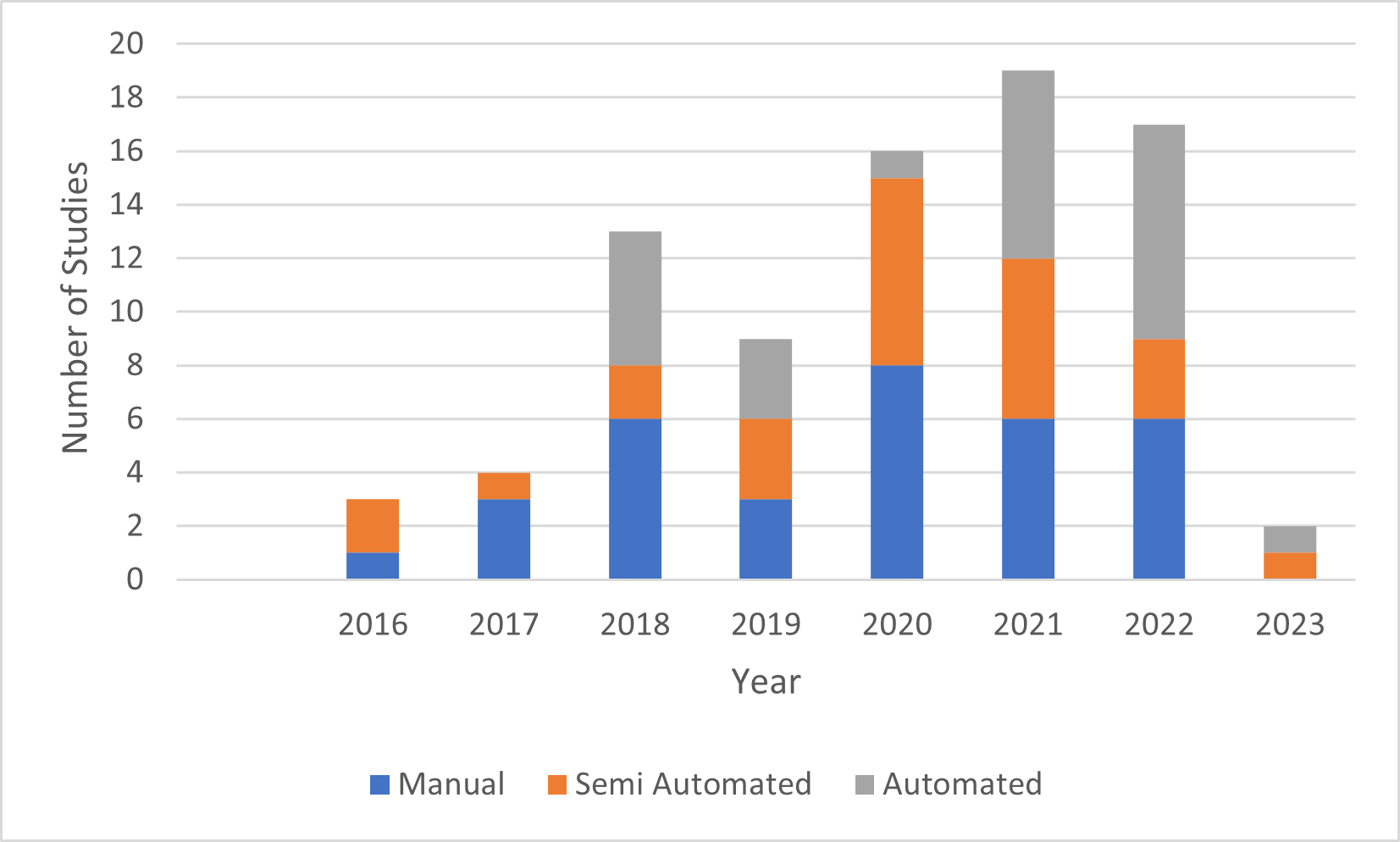} \label{fig:automationOverTime}}}%
\qquad
\subfloat[\centering Analysis type over time]{{\includegraphics[width=6.5cm]{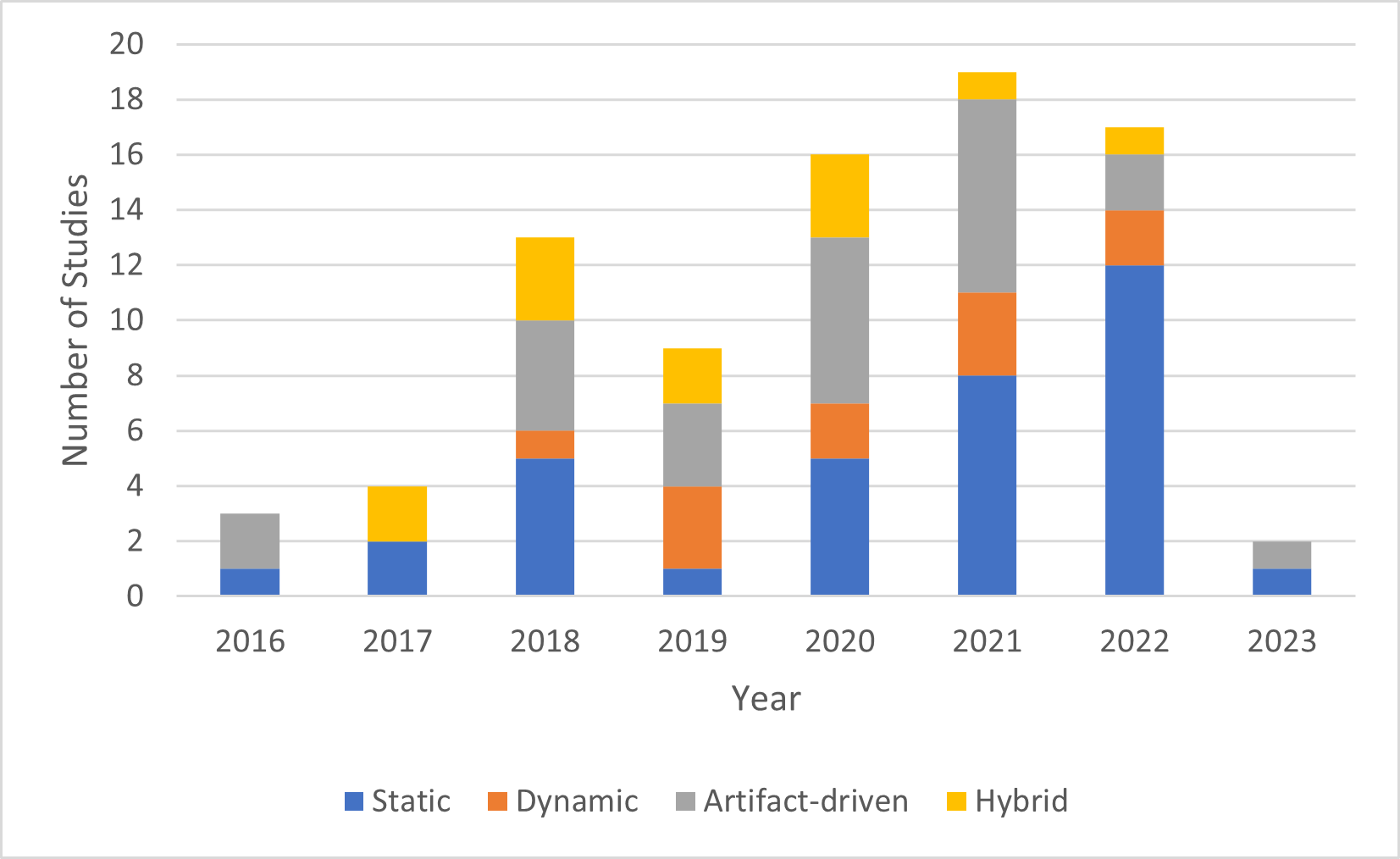} \label{fig:analysisTypeOverTime}}}%
\caption{Number of studies over time}%
    \label{fig:evaluationOverTime}%
\end{figure}

The first study on software systems reengineering into microservices was published in 2016. 
Manual, semi-automated, and automated techniques for migrating systems are discussed in the literature.
Manual techniques are completely human-oriented, whereas appropriate modeling, extraction, and visualization tools assist people during semi-automated system reengineering projects. 
In contrast, automatic techniques produce possible microservice recommendations from various inputs, e.g., source code, software logs, and software design artifacts. 
These recommendations can then form the basis for system reengineering.
\Cref{fig:automationOverTime}depicts the level of automation of the techniques in the surveyed studies over time. 
The majority of studies focused on the manual identification of services. 
However, a notable increase in semi-automated extraction techniques can be observed after 2018.
Moreover, attention to fully automated approaches increased after 2018.




The main identified approaches for decomposing software systems into microservices are static, dynamic, artifact-driven, and hybrid analysis. 
In static analysis, program source code, database schemata, and source code repository histories are used to provide insights into the system under study. 
By contrast, dynamic analysis considers execution time details like software system and server event logs, and runtime monitoring. 
The artifact-driven approaches are based on system artifacts like UML and data flow diagrams, architectural documents, use cases and user stories, ubiquitous language, and domain models.
Domain-driven design (DDD) and task-driven (functional-driven) design patterns are a subset of artificial-driven approaches. 
Finally, the hybrid approach can combine static, dynamic, and artifact-driven approaches.

\Cref{fig:analysisTypeOverTime} illustrates the microservice identification approaches published over time. 
Most of the existing studies are based on static system analysis. 
The artifact-driven analysis is the second most used technique for software systems reengineering. 
There was a notable increase in dynamic analyses during 2019 and 2020. 
Several hybrid techniques have been proposed over recent years.

\subsection{RQ2) What approaches are used to reengineer software systems into microservice-based systems, and how are reengineered systems evaluated?}

In this section, we discuss identified approaches for reengineering software systems. 
The algorithms and techniques practiced in literature and the availability of the tools for the reengineering are analyzed. 
Moreover, input/output data used to redesign software systems are reviewed.

\subsubsection{RQ2.1) What classes of approaches exist?}

The approaches used to analyze monolithic applications can be broadly classified into three main categories: static, dynamic, and artifact-driven analysis. 
An additional hybrid approach is identified, consolidating the main approaches.

The artifact-driven approaches use software artifacts like requirements, design diagrams, UML diagrams, data flow diagrams, business processes, use cases, user stories, domain models, and other design artifacts to identify bounded contexts for microservices. 
Each such bounded context implements a small, highly cohesive, loosely coupled behavior~\cite{MatiasMSBoundaries}. 
These contexts are then accepted as microservice candidates. 
The static analysis approaches are based on source code, database, and source code repository history analysis. 
The dependencies between classes, like inheritance, extended class relationships, the similarity between classes/database tables, and dependent commits in the repository, are observed in this approach. 
In contrast, the dynamic analysis approaches use runtime information to identify microservices. 
For example, they use runtime monitoring, execution time data correlations, and system-generated logs. 
Finally, the hybrid analysis techniques combine principles from the approaches discussed above.
Often, a hybrid approach results from extending a ``pure’’ technique with other analysis types. 
For example, a static analysis technique can borrow ideas of software log analysis to complement its microservice identification decisions.

\Cref{table:groupingDDD} details the five identified subcategories of the artifact-driven analysis (ADA) approaches.
The complete study list is provided in \Cref{appendix:studyList}. 
An ADA approach can use various system representations to analyze connected requirements, features, use cases, classes, and components.
Models representing the relationships between the classes/entities and UML diagrams are abstract representations of the software system. 
``Ubiquitous language'' refers to the terminology used in the software system. 
It is used to acquire all the terms used in the legacy system that represent business operations~\cite{JoselyneBNSystematicFrameworkOfApplicationMoernization}. 
Use cases determine the user interactions with the software system, whereas user stories explain system features. Architecture Description Language (ADL) and Unified Modeling Language (UML) visualize the software system's architecture.
These artifacts are used to identify the boundaries of services. 
In general, these artifacts are analyzed manually for the identification of microservice scopes and candidates.

A business process consists of activities coordinated in an organizational and technical environment to realize a business goal~\cite{WeskeBusinessPorcessManagement}. 
The dependencies between business processes implemented in a software system can be analyzed to identify data, structural, semantical, and control relationships. 
These dependencies are then represented as matrices and used to identify microservices using various clustering techniques.

A dataflow diagram (DFD) graphically illustrates the data processed by business functions or operations of a system in terms of inputs and outputs~\cite{ShanshanHZZCJQJZDataflowDrivenApproach}. 
It contains processes, data stores, data flow, and external entities. 
Highly expressive and detailed DFDs are used for microservice identification. 
Once a DFD is constructed, dependencies between processes and data stores are analyzed to identify microservices. 
This is done by constructing and analyzing dependency matrices to identify highly correlated processes/components or analyzing process and data store relationships using custom algorithms.

System responsibilities, requirements, features, and functionalities are also used to identify microservices~\cite{WeiFeatureTable}. 
The system functionalities are analyzed or divided into sub-tasks that cannot be divided further to identify the dependencies and relationships. 
Based on the identified relationships and dependencies, connected groups of functionalities are identified as microservices.

Finally, semantic analysis of domain artifacts relies on the analysis of further software artifacts~\cite{Daoud2020TowardsAA, Tusjunt}.
To identify microservices, these techniques conduct similarity calculations over the extracted vocabularies of system terms.
The extracted vocabulary is analyzed to identify related system entities and operations that can form microservices.

{
\small
\begin{table}[ht] 
\captionof{table}{Subcategories of ADA approaches}
\noindent\adjustbox{max width=\textwidth}{%
\begin{tabular}{|l|l|}
\hline
\textbf{Subcategory} & \textbf{Study IDs} \\ \hline
\multirow{2}{*}{\begin{tabular}[c]{@{}l@{}}Domain models, UML, ADL,  ubiquitous language, \\ use cases, user stories, and business reports\end{tabular}} & \multirow{2}{*}{\begin{tabular}[c]{@{}l@{}}1, 7, 8, 9, 40, 46, \\ 54, 75, 76\end{tabular}} \\
 &  \\ \hline
Business processes & 21, 59, 92 \\ \hline
Data flow diagrams & 48, 69 \\ \hline
\begin{tabular}[c]{@{}l@{}}System responsibilities, requirements, features, \\ and functions\end{tabular} & 41, 49, 65, 74, 77 \\ \hline
Semantic analysis of domain artifacts & 21, 73 \\ \hline
\end{tabular}}
\label{table:groupingDDD}
\end{table}
}
	
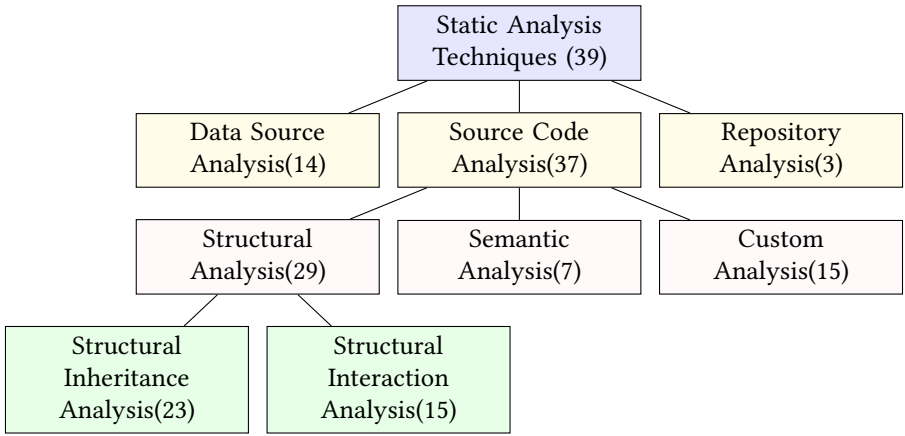
\begin{figure}[t]
\centering
\begin{forest}                      
for tree={
    grow=south,
    text width=3cm,%
    minimum height=2em,
    text centered,
    draw,
}
[Static Analysis Techniques (39), fill=blue!10
[Data Source\\ Analysis(14), fill=yellow!10] 
[Source Code\\ Analysis(37), fill=yellow!10
[Structural\\ Analysis(29), fill=pink!10
[Structural\\ Inheritance\\ Analysis(23), fill=green!10]
[Structural\\ Interaction\\ Analysis(15), fill=green!10]]
[Semantic Analysis(7), fill=pink!10]
[Custom\\ Analysis(15), fill=pink!10]] 
[Repository\\ Analysis(3), fill=yellow!10]]
\end{forest}
\caption{Classification of static analysis approaches}
\label{fig:classificationStatic}
\end{figure}

Static analysis for microservice identification is the most common approach discussed in the literature. 
Each analyzed static analysis technique uses one or several artifacts extracted from source code, database schemata and operations, and version control systems. 
The source code contains all the classes with system entities, core functions and business logic, communication APIs, and UI interfaces. 
Certain analyzed software systems reported that the persistent layer resides in the source code, including persistent entities and table mappings. 
The database consists of tables, table attributes, and relations. 
It was observed that the core business logic of legacy systems often resides in the database as stored procedures and functions. 
The version-control systems contain the histories of the source system files as collections of changes and commits. 
A commit is usually characterized by relevant change specifications or addressed requirements and an individual who implemented the changes. 
Several subcategories have been identified within static analysis techniques. 
They are reported in \Cref{fig:classificationStatic}.

The majority of existing Static Analysis Techniques are grounded in Source Code Analysis. 
Within techniques that analyze source code, Structural Analysis approaches use structural relationships between the classes and methods, Semantic Analysis methods examine the lexical relations within source code, while Custom Analysis techniques inspect source code from other perspectives, including API, annotations, and business logic.

The Structural Analysis approaches are further categorized into Structural Inheritance Analysis and Structural Interaction Analysis.
Structural inheritance analysis of source code explores the package, class, and method level dependencies and class inheritance hierarchies.
This analysis is often performed over an abstract syntax tree (AST) of source code, generating the system dependency graphs with classes/methods as vertices and dependencies as edges.  
Dependency graphs and ASTs are generated using existing tools.
The relevant studies are summarized in \Cref{table:staticFurtherClassification}.
Structural Interaction Analysis techniques examine the interconnections of classes and methods in source code. 
Three further subcategories have been identified within Structural Interaction Analysis techniques, as shown in \Cref{table:staticFurtherClassification}. 
The techniques from the first subcategory inspect APIs and other entry points like UI calls. 
A set of execution paths can be identified as a result of entry point analysis. 
Subpaths, interconnected paths, and data usage in the entry points were analyzed for microservice identification. 
The techniques that analyze call graphs and method calls form the second subcategory. 
Available open-source tools were used to generate the call graphs.
Call graphs can be context-sensitive or context-insensitive. 
Context-sensitive graphs ensure the different calls are annotated with different identifiers to distinguish different paths through the same sections of the code~\cite{NitinARKCargo}.
Finally, the techniques that fall into the third subcategory analyze object reference relationships, including the information flows that trigger the creation of instances of the objects.

Semantic Analysis techniques examine the similarity between the words (terminology) in source code and derive the co-related classes as possible microservice candidates.
This approach is also known as domain-related service decomposition. 
The core assumption for the techniques from this category is that related features use similar terminology at the implementation level. 
Further subclassification of the Semantic Analysis techniques is listed in \Cref{table:staticFurtherClassification}. 
Natural language processing (NLP) and information retrieval (IR) techniques are commonly used to extract semantic details. 
These techniques filter source code to exclude programming language keywords and space characters. 
Then, word tokenization, stop word removal, stemming, word enrichment using the synonyms from existing word dictionaries, and tf-idf calculations are performed. 
Instead of the entire source code, ASTs can be classified based on the similarity of terms. 
\Citet{BritoTopicModeling} use ASTs instead of source code to exclude the library dependencies to identify the terms of the system.
Like in the NLP-based approaches, stop word removal and stemming are applied to remove insignificant terms and reduce multiple variations of identical terms. 
After identifying the unique bag of words, topic modeling classifiers like Latent Dirichlet Allocation (LDA) and Seeded Latent Dirichlet Allocation (SLDA) are applied to group the lexical terms into clusters. 
These clusters are either directly identified as microservices or further processed using graph-based modeling. 
Moreover, unique term analyses have been conducted. 
They identify unique words in source code and create word frequency against each class file as a matrix to get the relatedness between two classes using cosine similarity calculation.

Moreover, multiple custom source code analysis techniques have been explored. 
Identification of business logic, persistent layer details extraction, reverse engineering the software system to extract the underlying architecture, language-specific annotation for extracting the system information, and API specification analysis like Open API are sub-categories of such custom analysis, as listed in \Cref{table:staticFurtherClassification}.
In addition, some techniques study the business logic of systems to identify the core business functions.
The identified business functions are then proposed to be grouped into microservices.
Persistent layer analysis relates to identifying entities associated with data sources and creating, reading, updating, and deleting (CURD) operations performed on data sources. 
Since the analysis is performed on the source code, it is also classified as a custom analysis technique. 
However, such analysis is often performed along with data source analysis. 
Reverse engineering is another custom analysis technique, extracting the underlying architecture using reverse engineering tools.
Extracted architecture is then used to apply other techniques, like dependency analysis, to obtain the related partitions of the system.
Programming language annotations, for instance, Java annotations like @EJB, @Controller, and @Entity, are used to identify certain code classes.
Finally, API specification techniques examine API documents, for example, API specification standards like OpenAPI\footnote{https://spec.openapis.org/oas/v3.1.0} to analyze their semantic similarity and use this information to inform microservice identification.

Enterprise systems use databases for data persistence. 
To identify microservices, it is essential to analyze persistent entities, like database tables, relevant to each microservice. 
In the literature, the need for a dedicated database per microservice pattern was highlighted \cite{TyszberwiczIdentifyMSFunctionDecomposition,NitinARKCargo}. 
Hence, studies for splitting the database layer have been conducted to be compatible with design principles. 
As a further categorization of data source analysis, database schema and table analysis, stored procedure analysis, SQL query mapping analysis, and topic modeling in tables were identified. 
The tables, table attributes, and relationships between the tables, like key constraints and triggers, were studied.

Due to performance issues and network overhead, business logic can be implemented in the database layer of a legacy system, for example, as stored procedures. 
Consequently, mapping between stored procedures and business functions constitutes another data source analysis technique. 
Moreover, this literature review revealed techniques for validating SQL queries and relevant business objects for microservice identification purposes.
Such techniques use information from SQL queries, relevant entities, and entity attributes accessed by SQL queries. 
Finally, topic modeling in database tables is another data source analysis technique. 
In such techniques, each table is considered a document, and table properties are considered to be the attributes of the document. 
The lexical similarity between the documents was used to categorize co-related tables that can be identified as highly cohesive partitions.

Evolutionary coupling analyzes source code repositories, mainly in the form of commit histories, to identify co-related classes in the change log. 
The consecutive commit analysis is a subcategory under evolutionary coupling, where change occurrences of multiple classes are analyzed to group them. 
Moreover, evolutionary coupling graphs are constructed by collecting the commits during a given period.
In such a graph, vertices represent classes, and an edge exists if two classes are associated in a single commit. 
This approach is known as logical coupling~\cite{MazlamiExtractionOfMicroservices}. 
The contributor coupling graph is the representation of developers who are involved in the changes in source code. 
Since well-organized teams are a main concern emphasized in the literature~\cite{MazlamiExtractionOfMicroservices}, this category is used to extract the system changes from the contributor's point of view. 
Similar to logical coupling, classes represented by vertices and edges exist if a particular developer is involved in the commit. 
Details on such studies are available in \Cref{table:staticFurtherClassification}.

{
\small
\begin{table}[t]
\captionof{table}{Static analysis approaches}
\noindent\adjustbox{max width=\textwidth}{%
\begin{tabular}{|l|l|l|}
\hline
\textbf{Category} & \textbf{Subcategory} & \textbf{Study IDs} \\ \hline
\multirow{2}{*}{Structural inheritance} & \begin{tabular}[c]{@{}l@{}}AST/system dependencies and\\ dependency graphs\end{tabular} & \begin{tabular}[c]{@{}l@{}}11, 32, 33, 43, 52, 53, 55, 57, \\ 66, 67, 80, 90, 91\end{tabular} \\ \cline{2-3} 
& \begin{tabular}[c]{@{}l@{}}Class hierarchy and inheritance/\\ subtype relationships\end{tabular} & \begin{tabular}[c]{@{}l@{}}27, 28, 29, 33, 37, 43, 44, 46, \\ 49, 52, 55, 63, 79, 89\end{tabular} \\ \hline
\multirow{3}{*}{Structural interaction} & APIs and entry points & 4, 29, 37, 39, 44 \\ \cline{2-3} 
& Call graphs and method calls & \begin{tabular}[c]{@{}l@{}}13, 28, 51, 53, 61, 66, 79, \\ 83, 89\end{tabular} \\ \cline{2-3} 
& Object reference relationships & 27, 53 \\ \hline
\multirow{3}{*}{Semantic analysis} & NLP- and IR-based analysis & 51, 66, 86 \\ \cline{2-3} 
& Topic modeling & 11, 55 \\ \cline{2-3} 
& Unique term analysis & 27, 28 \\ \hline
\multirow{5}{*}{Custom analysis} & Business logics & 4, 8, 91 \\ \cline{2-3} 
& \begin{tabular}[c]{@{}l@{}}Persistent layer entities/CRUD \\ operations\end{tabular} & 24, 25, 26, 27, 33, 61 \\ \cline{2-3} 
& Reverse engineering & 31, 79 \\ \cline{2-3} 
& Language annotations & 33, 79 \\ \cline{2-3} 
& API specification & 70, 85, 87 \\ \hline
\multirow{4}{*}{Data source analysis} & Schema and tables & \begin{tabular}[c]{@{}l@{}}4, 7, 27, 33, 40, 46, 49, \\ 53, 91\end{tabular} \\ \cline{2-3} 
& Stored procedures & 8 \\ \cline{2-3} 
& SQL queries to object mapping & 24, 25, 26 \\ \cline{2-3} 
& Topic modeling in DB tables & 58 \\ \hline
\multirow{3}{*}{\begin{tabular}[c]{@{}l@{}}Evolutionary coupling\\ analysis\end{tabular}} & Consecutive commit analysis & 32 \\ \cline{2-3} 
& Evolutionary coupling graphs & 51, 86 \\ \cline{2-3} 
& Contributor coupling graphs & 86 \\ \hline
\end{tabular}}
\label{table:staticFurtherClassification}
\end{table}
}

The final category of identified approaches is dynamic analysis. 
In such an approach, the software system is considered a black box. 
Based on the input provided, produced outputs are analyzed to identify frequent patterns and execution traces.
Three subcategories have been placed under dynamic analysis.
The first category relates to the analysis of server access log files. 
It is restricted to web applications. 
Web server access log files can be examined to identify the frequently invoking URIs. 
Server logs, e.g., Apache Tomcat\footnote{https://tomcat.apache.org/} and WildFly\footnote{https://www.wildfly.org/}, contain information on access URIs, request/response times, and response sizes. 
These logs are further analyzed based on URI frequency, response size, and response time to group the requests as possible candidate microservices. 
Most existing studies that perform dynamic analysis were based on system log analysis. 
Aspect-oriented programming (AOP) is a technique of log collection where an agent is embedded in source code to collect the logs based on the system entry points. 
In addition, one can use instrumented and deployed source code to collect logs.
Such logs were then provided as inputs to a process mining tool Disco\footnote{https://fluxicon.com/disco/} or custom analysis tools to get more insights into the source code.
Frequent execution traces, processes, and dependencies were observed by analyzing the collected logs.
Note that to collect such logs, user operations need to be performed on an installation of the system.
Multiple such strategies have been adopted in the literature, for instance, 
all use cases/user operations,
functional test execution,
unit test execution, and 
user behavior simulation.

Run time monitoring has been defined as another class of dynamic analysis approaches. 
In such an approach, the system is observed during execution time, and collected information is used for system reengineering.
Kieker\footnote{https://kieker-monitoring.net/}, Elastic APM\footnote{https://www.elastic.co/}, and dynatrace\footnote{https://www.dynatrace.com/} are the tools used for this purpose and mentioned in the literature.
\Cref{table:DynamicFurtherClassification} lists the categories and study IDs related to the dynamic analysis studies.
Furthermore, a hybrid approach can combine one or more artifact-driven, static, and dynamic analysis techniques. 
However, in a hybrid approach, there is often one dominant technique.
For example, static analysis can be conducted first and the extracted data complemented with dynamic analysis details for further analysis~\cite{MatiasMSBoundaries, Alwis, Alwis2018FunctionalSplitting}, or an artifact-driven analysis is conducted while static analysis information supports further analysis~\cite{Krause}.

{
\small
\begin{table}[t]
\captionof{table}{Dynamic analysis approaches}
\noindent\adjustbox{max width=\textwidth}{%
    \begin{tabular}{|l|l|}
    \hline
    \textbf{Category} & \textbf{Study IDs} \\ \hline
    Server access logs & 5, 6 \\ \hline
    \begin{tabular}[c]{@{}l@{}}System logs (AOP, instrumented deployment,\\ user operations and test executions)\end{tabular} & \begin{tabular}[c]{@{}l@{}}2, 24, 25, 26, 39, 42, 43,  \\ 57, 72, 83, 88\end{tabular} \\ \hline
    Run time monitoring & 13, 39, 46, 52, 88 \\ \hline
    \end{tabular}}
\label{table:DynamicFurtherClassification}
\end{table} 
}

\subsubsection{RQ2.2) What tools exist and which level of automation do they support?}

In the existing studies, two types of tools have been identified: tools/prototypes developed during the studies of microservice reengineering (in line with the concept in the study) and existing tools to support different stages of the reengineering process, e.g., call graph generation and log analysis.
In certain studies, source code is made freely available for use. 
Manual, partially automated, and automated approaches are used in literature for microservice extraction. 
The existing migration frameworks and their source code/tools with the level of automation are listed in \Cref{table:automationTool}. 
Tools that provide microservice recommendations based on primary inputs, like source code, log files, and system artifacts, are considered automated. 
The studies with tools involved in different stages of the migration process, like data extraction and system modeling, are categorized as semi-automated.

\begin{table}[t]
\captionof{table}{Tools and automation; level of automation: automated (A) and partially automated (PA)} 
\noindent\adjustbox{max width=\textwidth}{%
    \begin{tabular}{|llll|}
    \hline
    \multicolumn{1}{|l|}{\textbf{Study IDs}} & \multicolumn{1}{l|}{\textbf{Study details}} & \multicolumn{1}{l|}{\textbf{\begin{tabular}[c]{@{}l@{}}Level of \\ automation\end{tabular}}} & \multicolumn{1}{l|}{\textbf{Available artifacts}} \\ \hline
    \multicolumn{1}{|l|}{1} & \multicolumn{1}{l|}{\begin{tabular}[c]{@{}l@{}}Service Cutter: A Systmatic Approach \\ to service decomposition\end{tabular}} & \multicolumn{1}{l|}{PA} & https://github.com/ServiceCutter/ServiceCutter \\ \hline
    \multicolumn{1}{|l|}{11} & \multicolumn{1}{l|}{\begin{tabular}[c]{@{}l@{}}Identification of microservices from monolithic \\ applications through topic modelling\end{tabular}} & \multicolumn{1}{l|}{A} & https://github.com/miguelfbrito/microservice-identification \\ \hline
    \multicolumn{1}{|l|}{24} & \multicolumn{1}{l|}{\begin{tabular}[c]{@{}l@{}}Discovering Microservices in Enterprise Systems\\  Using a Business Object Containment Heuristic\end{tabular}} & \multicolumn{1}{l|}{A} & https://github.com/AnuruddhaDeAlwis/NSGAII \\ \hline
    \multicolumn{1}{|l|}{25} & \multicolumn{1}{l|}{\begin{tabular}[c]{@{}l@{}}Function-Splitting Heuristics for Discovery of \\ Microservices in Enterprise Systems\end{tabular}} & \multicolumn{1}{l|}{A} & https://github.com/AnuruddhaDeAlwis/Subtype \\ \hline
    \multicolumn{1}{|l|}{26} & \multicolumn{1}{l|}{\begin{tabular}[c]{@{}l@{}}Availability and Scalability Optimized Microservice \\ Discovery from Enterprise Systems\end{tabular}} & \multicolumn{1}{l|}{PA} & \begin{tabular}[c]{@{}l@{}}https://github.com/AnuruddhaDeAlwis/\\ NSGAIIFOROptimization\end{tabular} \\ \hline
    \multicolumn{1}{|l|}{29} & \multicolumn{1}{l|}{\begin{tabular}[c]{@{}l@{}}Graph Neural Network to Dilute Outliers for \\ Refactoring Monolith Application (CO-GCN)\end{tabular}} & \multicolumn{1}{l|}{PA} & https://github.com/utkd/cogcn \\ \hline
    \multicolumn{1}{|l|}{37} & \multicolumn{1}{l|}{Business process extraction using static analysis} & \multicolumn{1}{l|}{PA} & https://github.com/Rofiqul-Islam/logparser \\ \hline
    \multicolumn{1}{|l|}{39} & \multicolumn{1}{l|}{\begin{tabular}[c]{@{}l@{}}Service Candidate Identification from Monolithic \\ Systems Based on Execution Traces (FoSCI)\end{tabular}} & \multicolumn{1}{l|}{PA} & https://github.com/wj86/FoSCI/releases \\ \hline
    \multicolumn{1}{|l|}{42} & \multicolumn{1}{l|}{\begin{tabular}[c]{@{}l@{}}Mono2Micro: A Practical and Effective Tool for Decomposing \\ Monolithic Java Applications to Microservices\end{tabular}} & \multicolumn{1}{l|}{A} & https://www.ibm.com/cloud/mono2micro \\ \hline
    \multicolumn{1}{|l|}{51} & \multicolumn{1}{l|}{\begin{tabular}[c]{@{}l@{}}Steinmetz: Toward automatic decomposition of monolithic\\  software into microservices\end{tabular}} & \multicolumn{1}{l|}{PA} & \begin{tabular}[c]{@{}l@{}}https://github.com/loehnertz/Steinmetz/\\ https://github.com/loehnertz/semantic-coupling/\end{tabular} \\ \hline
    \multicolumn{1}{|l|}{52} & \multicolumn{1}{l|}{\begin{tabular}[c]{@{}l@{}}Determining Microservice Boundaries: A Case Study\\  Using Static and Dynamic Software Analysis\end{tabular}} & \multicolumn{1}{l|}{PA} & https://github.com/tiagoCMatias/monoBreaker \\ \hline
    \multicolumn{1}{|l|}{55} & \multicolumn{1}{l|}{\begin{tabular}[c]{@{}l@{}}Tool Support for the Migration to Microservice \\ Architecture: An Industrial Case Study\end{tabular}} & \multicolumn{1}{l|}{A} & https://essere.disco.unimib.it/wiki/arcan/ \\ \hline
    \multicolumn{1}{|l|}{58} & \multicolumn{1}{l|}{\begin{tabular}[c]{@{}l@{}}Towards Migrating Legacy Software Systems to \\ Microservice-based Architectures: a Data-Centric Process \\ for Microservice Identification\end{tabular}} & \multicolumn{1}{l|}{PA} & https://bit.ly/31ySia7 \\ \hline
    \multicolumn{1}{|l|}{61} & \multicolumn{1}{l|}{\begin{tabular}[c]{@{}l@{}}Microservices Identification in Monolith Systems: Functionality \\ Redesign Complexity and Evaluation of Similarity Measures\end{tabular}} & \multicolumn{1}{l|}{A} & https://github.com/socialsoftware/mono2micro/tree/master \\ \hline
    \multicolumn{1}{|l|}{70} & \multicolumn{1}{l|}{\begin{tabular}[c]{@{}l@{}}Expert system for automatic microservices identification using \\ API similarity graph\end{tabular}} & \multicolumn{1}{l|}{A} & https://github.com/HduDBSI/MsDecomposer \\ \hline
    \multicolumn{1}{|l|}{77} & \multicolumn{1}{l|}{\begin{tabular}[c]{@{}l@{}}A Feature Table approach to decomposing monolithic applications \\ into microservices\end{tabular}} & \multicolumn{1}{l|}{PA} & https://github.com/RLLDLBF/FeatureTable/ \\ \hline
    \multicolumn{1}{|l|}{79} & \multicolumn{1}{l|}{Leveraging the Layered Architecture for Microservice Recovery} & \multicolumn{1}{l|}{PA} & \begin{tabular}[c]{@{}l@{}}https://gitlab.com/LeveragingInternalArchitecture/\\ IdentificationApproach\end{tabular} \\ \hline
    \multicolumn{1}{|l|}{86} & \multicolumn{1}{l|}{\begin{tabular}[c]{@{}l@{}}Extraction of Microservices from Monolithic Software Architectures\\ (MEM)\end{tabular}} & \multicolumn{1}{l|}{PA} & \begin{tabular}[c]{@{}l@{}}https://github.com/gmazlami/microserviceExtraction-backend\\ \\ https://github.com/gmazlami/microserviceExtraction-frontend\end{tabular} \\ \hline
    \multicolumn{1}{|l|}{89} & \multicolumn{1}{l|}{\begin{tabular}[c]{@{}l@{}}From legacy to microservices: A type‐based approach \\ for microservices identification using machine learning and \\ semantic analysis (microminer)\end{tabular}} & \multicolumn{1}{l|}{A} & \begin{tabular}[c]{@{}l@{}}https://drive.google.com/drive/folders/\\ 1TQaS8etLr-32d0RXwC1Le-IOMVaDBcSS?usp=sharing\end{tabular} \\ \hline
    \end{tabular}}
\label{table:automationTool}
\end{table}

Multiple categories of tools are available based on the approaches used to examine the monolithic system. 
There are tools for the static analysis of software systems, database administration, runtime monitoring, visualization, architectural validation, and load simulations. These tools, technologies used, and respective study IDs are listed in \Cref{table:toolDetailsTable}.
Moreover, a comparison between existing tools utilized to extract microservices has been made in a separate study by \citet{toolAnalyse}. 
\citet{EasyAPM} used their tool EasyAPM to record the data operation and parameter information through the instrument on the JDBC and data access class libraries. 
Other supportive tools used for testing, clustering, and other specific purposes are listed in \Cref{table:additionalTools}.

\begin{table}[t]
\captionof{table}{Tools for monolithic system analysis}
\noindent\adjustbox{max width=\textwidth}{%
    \begin{tabular}{|l|l|l|l|l|}
    \hline
    \textbf{Purpose} & \textbf{Tool} & \textbf{Details} & \textbf{Technology} & \textbf{Study IDs} \\ \hline
    \multirow{11}{*}{\begin{tabular}[c]{@{}l@{}}Static Analysis \\ (Source Code)\end{tabular}} & Java Call Graph & \begin{tabular}[c]{@{}l@{}}Read the jar file to collect the method \\ calling sequence. Dynamic analysis exists\\ but is used in static context only.\end{tabular} & Jave & 13 \\ \cline{2-5} 
     & \begin{tabular}[c]{@{}l@{}}Java Parser Symbol \\ Resolver\end{tabular} & Structural dependency extraction. & Java & 11 \\ \cline{2-5} 
     & Mondrian & Static source code analysis. & PHP & 27, 28 \\ \cline{2-5} 
     & Java Parser & Constructs abstract syntax tree. & Java & 44 \\ \cline{2-5} 
     & WALA & \begin{tabular}[c]{@{}l@{}}Project class hierarchy analysis and call\\ graph generation.\end{tabular} & Java, Java Script & 53, 57 \\ \cline{2-5} 
     & Soot & \begin{tabular}[c]{@{}l@{}}Models source code. Analyze, instrument, \\ optimize, and visualize applications.\end{tabular} & Java, Android & 29, 53 \\ \cline{2-5} 
     & Doop \& Datalog & \begin{tabular}[c]{@{}l@{}}Static analysis of source code with Datalog\\ engine.\end{tabular} & Java, Android & 53 \\ \cline{2-5} 
     & JackEE & \begin{tabular}[c]{@{}l@{}}Static analysis of Java Web applications; \\ enterprise framework support.\end{tabular} & JEE applications & 53 \\ \cline{2-5} 
     & Spoon & \begin{tabular}[c]{@{}l@{}}Source code analysis tool. Parse source\\ code into abstract syntax tree.\end{tabular} & Java & 61 \\ \cline{2-5} 
     & Structure 101 & \begin{tabular}[c]{@{}l@{}}Architecture validation and maintainance \\ tool with visualizing the structure from \\ source code.\end{tabular} & \begin{tabular}[c]{@{}l@{}}Java, .Net, C/C++\\ and more\end{tabular} & 2, 46, 71 \\ \cline{2-5} 
     & \begin{tabular}[c]{@{}l@{}}Sonagraph\\ Architect\end{tabular} & \begin{tabular}[c]{@{}l@{}}Static analysis tool with architecture check,\\ code duplicate check, virtual refactorings,\\ breakup cyclic dependencies and comparison\\ with previous versions. Supports git\\ repository mining.\end{tabular} & \begin{tabular}[c]{@{}l@{}}C\#, C/C++, Java,\\ Python 3\end{tabular} & 77 \\ \hline
    \multirow{3}{*}{\begin{tabular}[c]{@{}l@{}}Static Analysis\\ (Data Base)\end{tabular}} & SchemaSpy & \begin{tabular}[c]{@{}l@{}}Generate Web based visual representation \\ by analysing database meta data.\end{tabular} & Java based tool & 2, 71, 72 \\ \cline{2-5} 
     & DBeaver & \begin{tabular}[c]{@{}l@{}}Database administration tool and\\ analysis of database schema.\end{tabular} & \begin{tabular}[c]{@{}l@{}}MySQL, Maria DB,\\ PostgreSQL, SQLite\end{tabular} & 46 \\ \cline{2-5} 
     & JSQLParser & \begin{tabular}[c]{@{}l@{}}Parse SQL statements and translate them to \\ hierarchies of Java classes.\end{tabular} & Java \& SQL & 61 \\ \hline
    \multirow{7}{*}{Dynamic Analysis} & Kieker & \begin{tabular}[c]{@{}l@{}}Monitoring and analyzing runtime behavior\\ of software systems.\end{tabular} & \begin{tabular}[c]{@{}l@{}}Java, .Net, C language,\\ VB\end{tabular} & 13, 39, 88 \\ \cline{2-5} 
     & Elastic APM & \begin{tabular}[c]{@{}l@{}}Application performance monitoring system.\\ Supports realtime monitoring, incoming \\ request/response performance, database\\ queries, cache invocations, external calls.\end{tabular} & \begin{tabular}[c]{@{}l@{}}Java-based, Web, data\\ access frameworks, \\ application servers, \\ messaging frameworks,\\ AWS\end{tabular} & 2, 72 \\ \cline{2-5} 
     & Disco & \begin{tabular}[c]{@{}l@{}}Process mining tool with event log analysis to\\ identify call graphs and automated process\\ discovery.\end{tabular} & \begin{tabular}[c]{@{}l@{}}Log files of software\\ systems\end{tabular} & \begin{tabular}[c]{@{}l@{}}2, 24, 25, \\ 26\end{tabular} \\ \cline{2-5} 
     & Dynatrace & \begin{tabular}[c]{@{}l@{}}Monitoring platform for cloud environment\\ with log management and analytics, business\\ process \& infrastructure monitoring and end\\ to end observability of applications.\end{tabular} & \begin{tabular}[c]{@{}l@{}}Java, Python, \\ cloud technologies\end{tabular} & 72 \\ \cline{2-5} 
     & Datadog & \begin{tabular}[c]{@{}l@{}}AI powered code level distributed tracing\\ from browser and mobile apps to backend\\ and DB and log analysis.\end{tabular} & \begin{tabular}[c]{@{}l@{}}Java, Python, cloud \\ technologies and more\end{tabular} & 72 \\ \cline{2-5} 
     & ExplorViz & \begin{tabular}[c]{@{}l@{}}Runtime monitoring and visualization of \\ software.\end{tabular} & \begin{tabular}[c]{@{}l@{}}Applied to Java based\\ system\end{tabular} & 46 \\ \cline{2-5} 
     & django-silk & \begin{tabular}[c]{@{}l@{}}Profiling and inspection tool for the django \\ framework. Analyze HTTP requests and \\ database queries.\end{tabular} & \begin{tabular}[c]{@{}l@{}}Python django\\ framework based tools\end{tabular} & 52 \\ \hline
    \end{tabular}}
\label{table:toolDetailsTable}
\end{table}

\begin{table}[h]
\captionof{table}{Additional tools used for system analysis}
\noindent\adjustbox{max width=\textwidth}{%
    \begin{tabular}{|l|l|l|l|l|}
    \hline
    \textbf{Purpose} & \textbf{Tool} & \textbf{Details} & \textbf{Technology} & \textbf{Study IDs} \\ \hline
    \multirow{2}{*}{Testing} & Jmeter & Load simulation & Java & 6, 7 \\ \cline{2-5} 
     & Gatling & Stress testing & Java, Kotlin, Scala & 16 \\ \hline
    Reverse engineering & MoDisco & \begin{tabular}[c]{@{}l@{}}Model driven reverse \\ engineering framework\end{tabular} & Java, JEE, XML & 31 \\ \hline
    Topic modeling & GuidedLDA & \begin{tabular}[c]{@{}l@{}}Topic modeling using\\ latent Dirichlet allocation\end{tabular} & Python & 55 \\ \hline
    Clustering & SciPy & \begin{tabular}[c]{@{}l@{}}Tool for hierarchical clustering\\ and generate dendograms\end{tabular} & Python & 61 \\ \hline
    Optimization algorithm & Jmetal & \begin{tabular}[c]{@{}l@{}}Multi-objective optimization\\ algorithm\end{tabular} & Java & 82 \\ \hline
    Document enrichmnet & \begin{tabular}[c]{@{}l@{}}WordWeb,\\ WordNet\end{tabular} & Identify synonyms for topic modeling & Word dictionary & 58 \\ \hline
    Lines of code count & CLOC & \begin{tabular}[c]{@{}l@{}}Blank, comment and physical lines \\ counting.\end{tabular} & \begin{tabular}[c]{@{}l@{}}Java, C, Python\\ and more\end{tabular} & 12 \\ \hline
    \end{tabular}}
\label{table:additionalTools}
\end{table}

\subsubsection{RQ2.3) Which techniques/algorithms are used?}

This section discusses techniques and algorithms used during the reengineering of software systems.
Identified techniques can be broadly categorized into two types: system modeling techniques and microservice extraction techniques. 
The system modeling techniques provide an abstract representation of the software system, whereas extraction techniques identify the boundaries of the microservices within the modeled system. 
Identified system modeling techniques and relevant studies are listed in \Cref{table:modelingTechniques}.
 
{
\small
\begin{table}[t]
\captionof{table}{System modeling techniques}
\noindent\adjustbox{max width=\textwidth}{%
    \begin{tabular}{|l|l|}
    \hline
    \textbf{Technique} & \multicolumn{1}{c|}{\textbf{Study IDs}} \\ \hline
    Graph-based modeling & \begin{tabular}[c]{@{}l@{}}1, 11, 13, 21, 24, 25, 26, 29, 31, 32, 37, 43, 44, \\ 49, 51, 52, 53, 55, 57, 61,  70, 79, 82, 83, \\ 86, 89, 90, 91\end{tabular} \\ \hline
    Matrix/table based modeling & \begin{tabular}[c]{@{}l@{}}7, 8, 25, 27, 28, 33, 50, 59, 65, 66, 69, 75, 77, \\ 80, 85, 87, 92\end{tabular} \\ \hline
    URI frequency-based modeling & 5, 6 \\ \hline
    Domain element-based modeling & 33, 40, 41, 46, 48, 54 \\ \hline
    Execution traces modeling & 2, 25, 39, 42, 72, 88 \\ \hline
    Semantic modeling & 55, 58, 73, 74, 85, 87 \\ \hline
    \end{tabular}}
\label{table:modelingTechniques}
\end{table}
}

The first approach to modeling legacy systems uses graphs, the graph-based modeling technique. 
It is an extensively practiced technique for modeling software systems.
The vertices in the graphs can be components, system entities, classes, methods, business processes, entry points, execution traces, database tables, and system functionalities. 
Edges can be either weighted or non-weighted. 
Undirected weighted edges are frequently used for the system graphs. 
The existence of an edge and its weight are based on the strength of the relationship between two vertices.
Structural relationship graphs are constructed based on the number of dependencies, method calls, and coupling scores.
Dependencies and method calls are directly derived from ASTs, call graphs, and dependency graphs.
Moreover, structural relationships can be prioritized by assigning weights based on their types, e.g., generalization, aggregation, implementation, association, instantiation, and method invocation~\cite{AbdellatifTypeBasedApproachML}. 

Static dependency graphs are generated by static analysis tools. 
Subsequently, the coupling scores are often calculated manually based on the pre-defined parameters. 
Depending upon the four categories of cohesiveness, compatibility, constraints, and communication, 16 coupling criteria have been defined by \citet{GyselServiceCutter}. 
A priority and score can be defined for each criterion that contributes to the final edge weight of the graph. 
Semantic similarity-based graphs are based on tf-idf (term frequency-inverse document frequency) or topic-modeling. 
Once the tf-idf is calculated, a vector with the frequency of each word distribution in the class is obtained.
The cosine similarity between two vectors is calculated, capturing the degree of similarity between two data points.
A high degree of similarity defines the closely related classes.

For the topic modeling graphs, Probabilistic Latent Semantic Analysis (PLSA), Latent Dirichlet Allocation (LDA), Latent Semantic Analysis (LSA), and Non-negative Matrix Factorisation (NMF) are four major classes of algorithms. 
High coherence and fewer overlaps between the clusters were identified for LDA~\cite{PoojaPTopicModelingReview}. 
Therefore, Latent Dirichlet Allocation (LDA) and Seeded Latent Dirichlet Allocation (SLDA) are commonly used to identify the topic distribution within the source code. 
LDA is a probabilistic topic model. 
It is based on the distribution of the topics. 
It is an unsupervised model, whereas SLDA is a semi-supervised variant of LDA.
SLDA accepts the list of keywords as input that will stimulate the expected topics. 
Once the clusters are identified, cosine similarities between the clusters are calculated to define the edge weights in the graph representation. 

Graph edges based on dynamic analysis are defined based on the run time frequencies of method invocations and execution traces.
Evolutionary coupling graph edges are based on correlated classes in the commits and contributors that developed the classes. 
Examples of graph models include classes/components as vertices and topic modeling strengths as the edge weights, domain entities as vertices and coupling scores as the edge weight, call graphs with classes/methods as the vertices and execution calls and frequency as the edges, system entities and entry points as vertices and edges for method calls, system classes/entities against evolutionary coupling (connected classes in revision history) as edges, classes as vertices and edges based on contributors involved for development, classes/methods as vertices and runtime invoking relationships and frequencies as edges, and reverse engineering tool generated architectural graphs.
 
Matrix/table-based modeling represents a software system as a mapping of its attributes and components captured in a matrix with the number of occurrences as entries to classify the co-related attributes further.
Once matrices are constructed, similarity measures are used to identify related components that can define microservices. 
Either classes, methods, database tables/entities, use cases, sub-graphs, micro-tasks (tasks that cannot be decomposed further), or business processes are used in the computations of the frequencies of executions, sub-type/reference relationships, coupling, and cohesion values to determine the relationship between elements.
Semantic similarity analysis uses classes against unique word matrices. 
Then, cosine similarity determines the semantic similarity between the classes.
Example matrix/table-based modeling techniques include use-case-to-use-case similarity and use-case-to-database-entity similarity matrices, subgraph similarity matrices, class-to-database-object matrices, class-subtype (subtype relationships between classes) matrices, class-reference-type matrices, micro-tasks-to-data-object matrices, business process dependency matrices, structural similarity matrices (structural relationships between classes in a matrix format), conceptual similarity matrices (semantic similarity between classes in a matrix format), read/write operations between primitive types (further non-decomposable functions) and data storage, user story coupling and cohesion matrices, BPMN structural and data dependency matrices, feature tables, and use case to business process mapping tables.

URI-based modeling is an approach to modeling web applications. 
Web applications operate on a request/response base, where features are requested via URI calls, and responses are redirected to the relevant clients. 
Application servers like Tomcat and JBoss record logs with the request/response details. 
These details are used to infer models of the applications and identify the frequent URI calls that can be isolated as separate services for better performance. 
Mean request/response time (MRRT) and response size are used as indicators of network overhead and resource utilization.

Domain element-based modeling is another approach used to represent software systems. 
This approach uses data flow diagrams, UML diagrams, system capability models, and context maps to represent the software systems. 
This is a manual approach with detailed system diagrams with fine-grain information and capabilities that are analyzed to identify bounded contexts.

Execution trace modeling uses software logs to identify the actual methods/classes invoked during the runtime of the software systems. The collection of active execution traces defines the overall behavior of the system. In addition, inactive paths can be identified in the runtime traces analysis~\cite{KaliaMonoToMicro}. Multiple techniques have been used in the literature to investigate these execution traces. One approach is providing the software logs into the runtime trace analysis tool, e.g., DISCO process mining tool~\cite{TaibiProcessMining}. Tool-generated execution call graphs can be used to analyze and extract the co-related classes/methods manually~\cite{TaibiProcessMining} or programmatically identify the subtypes and common subgraphs~\cite{Alwis2018FunctionalSplitting}. Execution traces can be further modeled and reduced to identify functional atoms, which are coherent and minimal functional units~\cite{JinExecutionTracesFosci}, identify direct/indirect call patterns in execution traces~\cite{KaliaMonoToMicro}, and analyze class and method level execution traces based on system functionalities~\cite{Jin2018FoME}.

Semantic-based modeling is used to model the system based on linguistic information. 
Identifying system topics based on the application domain~\cite{Pigazzini2019ToolSF}, generating a vocabulary tree to illustrate the system terminology~\cite{Tusjunt}, and examining the system subject and operations to group the terms used in the API specifications~\cite{BaresiGD2017InterfaceAnalysis, AlDebagyPNewDecompositionMethodHyperParameter} have been done in semantic-based modeling.

\begin{figure}[t]	
\resizebox{\textwidth}{!}{%
    \begin{forest}
        for tree={
            child anchor=west,
            parent anchor=east,
            grow=east,
            text width=4cm,%
            draw,
            anchor=west,
            edge path={
                \noexpand\path[\forestoption{edge}]
                (.child anchor) -| +(-5pt,0) -- +(-5pt,0) |-
                (!u.parent anchor)\forestoption{edge label};
            },
        }
        [Extraction Techniques, fill=blue!10  
        [Custom Techniques, fill=yellow!10]
        [Genetic Algorithms, fill=yellow!10
        [Other Genetic Algorithms, fill=pink!10]
        [Multi-Objective\\ Optimization Algorithms, fill=pink!10]]
        [Neural Networks, fill=yellow!10]
        [Matrix/Table-Based\\ Extraction, fill=yellow!10
        [Custom Techniques, fill=pink!10]
        [Matrix/Table Clustering, fill=pink!10]]
        [Graph-Based Extraction, fill=yellow!10
        [Custom Clustering\\ Techniques, fill=pink!10]
        [Community Detection, fill=pink!10]
        [Clustering, fill=pink!10]]]
    \end{forest}
}
\caption{Microservice extraction techniques}
\label{fig:extractionTechniques}
\end{figure}
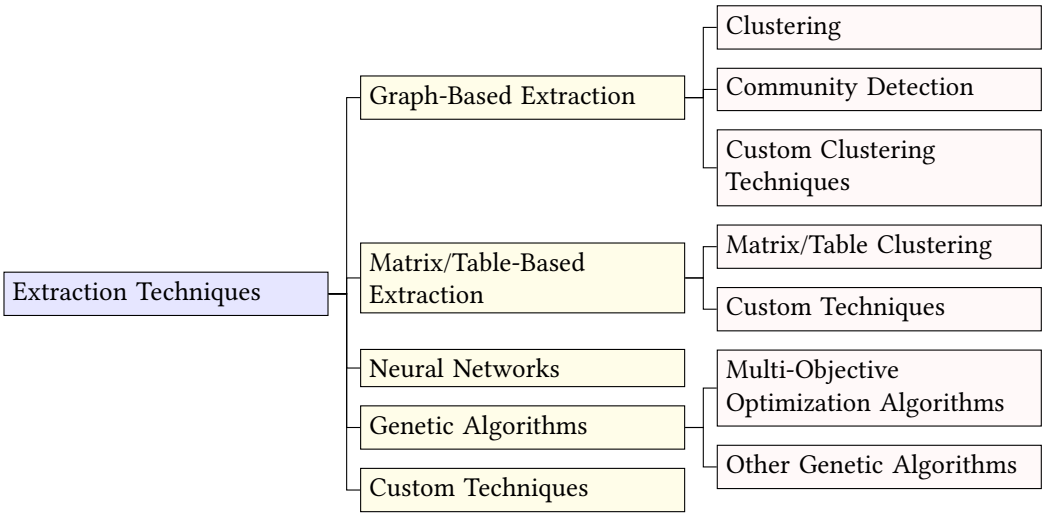

\Cref{fig:extractionTechniques} illustrates the identified microservice extraction techniques. 
After modeling the system using the aforementioned techniques, extraction is performed to identify the microservice candidates. 
Several studies identify microservices from system models, e.g., clusters identified from topic-based modeling, and tool-generated results after log analysis.

Graph-based extraction is the prominent technique based on graph-based modeling used to extract microservices. 
Hierarchical clustering is used when the number of clusters is not given as an input. 
By contrast, K-means clustering is used when prior knowledge of the number of desired clusters (microservices) is available. 
The advantage of parameterizing the number of clusters is the ability to analyze the service decomposition with any possible number of services. 
It can be used for a better understanding of the system and coupling between the parts of the system~\cite{GyselServiceCutter}. 
Two variations of hierarchical clustering that are used are agglomerative clustering and divisive clustering.
Agglomerative clustering starts with data points and iteratively generates the clusters, whereas divisive clustering starts with the complete dataset and splits it into clusters. 
Furthermore, temporo-spatial clustering is a variant of hierarchical clustering, and collaborative clustering is a variant of hierarchical agglomerative clustering. 
\Cref{fig:extractionTechniques} summarizes the studies.

Community detection studied in large-scale networks has been applied for microservice extraction from graph models. 
For instance, Girvan-Newman deterministic~\cite{NewmanGirvanClusteringAlgo} and Epidemic Label Propagation (ELP)~\cite{RaghavanNARKELPAlgo} non-deterministic algorithms were applied to discover microservices.
ELP algorithm takes in the number of clusters as an input parameter.
Louvain and fast community detection algorithms are based on maximizing modularity within a given network.
Louvain algorithm is an unsupervised algorithm. 
It is based on modularity maximization and does not require the number of communities or the size of the communities as input~\cite{BritoTopicModeling}.
Among the algorithms evaluated for microservice detection purposes are MCL, Walktrap, Clauset et al., Louvain, label propagation, infomap, and Chinese whispers~\cite{JakobOSteinmetz}.
It was observed that the Louvain algorithm performed the best to support the identification of microservices.

Hierarchical agglomerative clustering is used to analyze matrix/table-based models of systems. 
DBSCAN is a density-based clustering algorithm that aims to group elements that are densely packed in the search space and identify noisy elements that do not fit into any clusters using two concepts, which are neighborhood distance and the minimum number of elements in a neighborhood~\cite{SellamiDBSCAN}.

The Non-dominated Sorting Genetic Algorithm II (NSGA II) and Non-dominated Sorting Genetic Algorithm III (NSGA III) are multi-objective optimization algorithms. 
A multi-objective optimization algorithm aims to provide optimal solutions while achieving global optima when multiple conflicting objectives, e.g., coupling, cohesion, and modularity, are to be considered~\cite{Alwis}. 
Multiple studies have compared the performance of NSGA II and NSGA III~\cite{PranavaARNAComparisonOfNSGA, IshibuchiISYPerformanceComparisonOfNSGA}. 
It was pointed out that NSGA III does not consistently outperform NSGA II in microservice discovery.

Several custom extraction techniques have been identified in the literature. 
Manual and expert analysis are basic extraction techniques, with artifact-driven approaches being the most widely used manual microservice identification approaches.

{
\small
\begin{table}[t]
\captionof{table}{Microservice extraction techniques}
\noindent\adjustbox{max width=\textwidth}{%
\begin{tabular}{|l|l|l|}
\hline
\multicolumn{1}{|c|}{\textbf{Category}} & \multicolumn{1}{c|}{\textbf{Algorithm}} & \multicolumn{1}{c|}{\textbf{Study IDs}} \\ \hline
\multirow{8}{*}{Clustering algorithms} & Hierarchical clustering & 61, 79, 88 \\ \cline{2-3} 
 & K-means hierarchical clustering & 57, 70 \\ \cline{2-3} 
 & Collaborative clustering & 21 \\ \cline{2-3} 
 & Temporo spacial clustering & 43 \\ \cline{2-3} 
 & SArF clustering algorithm & 44 \\ \cline{2-3} 
 & MCL algorithm & 51 \\ \cline{2-3} 
 & Chinese Whispers & 51 \\ \cline{2-3} 
 & Affinity propagation algorithm & 87, 90 \\ \hline
\multirow{5}{*}{\begin{tabular}[c]{@{}l@{}}Community detection \\ algorithms\end{tabular}} & Girvan-Newman algorithm & 1, 52 \\ \cline{2-3} 
 & Label Propagation & 1, 51, 53 \\ \cline{2-3} 
 & Leiden algorithm & 13, 80 \\ \cline{2-3} 
 & Louvain community detection algorithm & 11, 49, 89 \\ \cline{2-3} 
 & Fast community algorithm & 32 \\ \hline
\multirow{2}{*}{Custom clustering} & Minimum spanning tree based detection & 86 \\ \cline{2-3} 
 & EJB based clustering & 31 \\ \hline
\multirow{3}{*}{Table/matrix clustering} & K-means clustering & \begin{tabular}[c]{@{}l@{}}5, 6, 27, 28, \\ 58, 59\end{tabular} \\ \cline{2-3} 
 & Hierarchical clustering & 7, 42, 65 \\ \cline{2-3} 
 & DBSCAN & 66 \\ \hline
\multirow{2}{*}{\begin{tabular}[c]{@{}l@{}}Custom matrix-based \\ identification\end{tabular}} & Matrix based ranking & 2, 72 \\ \cline{2-3} 
 & Subgraph similarity matrix & 25 \\ \hline
\multirow{3}{*}{Genetic algorithms} & NSGA II (Multi-objective optimization) & 24, 26, 39 \\ \cline{2-3} 
 & NSGA III (Multi-objective optimization) & 82, 83 \\ \cline{2-3} 
 & Other genetic algorithms & 75,92 \\ \hline
Neural networks &  & 29 \\ \hline
\multirow{6}{*}{Custom extraction} & Expert analysis & 55 \\ \cline{2-3} 
 & \begin{tabular}[c]{@{}l@{}}Semantic and domain element\\ analysis\end{tabular} & \begin{tabular}[c]{@{}l@{}}40, 41, 46, \\ 73, 74, 85\end{tabular} \\ \cline{2-3} 
 & DFD \& UML mapping analysis & 48, 54, 69 \\ \cline{2-3} 
 & Feature modeled system & 8 \\ \cline{2-3} 
 & Dependency and call relation analysis & 33, 91 \\ \cline{2-3} 
 & Rule based identification & 77 \\ \hline
\end{tabular}}
\label{table:extractionTechniques}
\end{table}
}

\subsubsection{RQ2.4) How data are used?}

Next, we discuss the inputs and outputs of existing approaches for reengineering software systems into microservices.

The artifact-driven approaches use artifacts like UML diagrams, Data Flow Diagrams (DFD), use cases, user stories, and architectural documents as inputs.
Note that most existing artifact-driven reengineering approaches are manual. 
At the same time, several existing studies convert system artifacts to computer-readable formats or intermediate representations and use them to identify candidate microservices.
After such (semi-)automatic identification, recommended microservice candidates are delivered as output.
Such outputs can have visual representation or be given as clusters of elements.
\Cref{table:inputOutputDDD} summarizes formats of inputs and outputs used by the artifact-driven approaches.
This table only covers studies with clearly identified input and output details. 

\begin{table}[h]
\captionof{table}{Inputs and outputs of artifact-driven approaches}
\noindent\adjustbox{max width=\textwidth}{%
    \begin{tabular}{|l|l|l|}
    \hline
    \textbf{Study ID} & \textbf{Input and/or intermediate representation} & \textbf{Output} \\ \hline
    1 & \begin{tabular}[c]{@{}l@{}}Json-based representation of SSA to identify nano entities to \\ generate a graph with coupling as the edge weights\end{tabular} & \begin{tabular}[c]{@{}l@{}}Entities grouped into clusters to represent \\ microservices (service cuts).\end{tabular} \\ \hline
    7 & \begin{tabular}[c]{@{}l@{}}Component-attribute data set. Use case to use case (U-to-U)  and \\ use case to DB entity (U-to-DB) relationship matrix to generate the \\ similarity matrix between the use cases.\end{tabular} & \begin{tabular}[c]{@{}l@{}}Candidate microservices with use \\ case grouping.\end{tabular} \\ \hline
    8 & \begin{tabular}[c]{@{}l@{}}Microservice Discovery Table (MDT) with  requirements, features, \\ and stored procedures mapping.\end{tabular} & \begin{tabular}[c]{@{}l@{}}MDT augmented with information on microservices, \\ entities, and rules.\end{tabular} \\ \hline
    21 & \begin{tabular}[c]{@{}l@{}}Business processes and dependencies (control, semantic, data, \\ and organizational) and dependency score matrix.\end{tabular} & Groups of cohesive activities. \\ \hline
    40 & \begin{tabular}[c]{@{}l@{}}Ubiquitous language, business operations, data operations, \\ domain models, database schema, design documents.\end{tabular} & \begin{tabular}[c]{@{}l@{}}Bounded contexts obtained after DDD pattern \\ analysis, business operation and data \\ dependency analysis.\end{tabular} \\ \hline
    41 & System responsibilities obtained and ubiquitous language. & Identified candidate microserive boundaries. \\ \hline
    48 & Data flow diagram (DFD) of the system. & \begin{tabular}[c]{@{}l@{}}Set of decomposable DFDs and grouping \\ of DFDs as microservices.\end{tabular} \\ \hline
    50 & \begin{tabular}[c]{@{}l@{}}System functionalities -- Mapping between business \\ requirements and system services. Task dependency matrix\\ for clustering.\end{tabular} & \begin{tabular}[c]{@{}l@{}}Task decomposition as clusters to represent\\  microservices.\end{tabular} \\ \hline
    54 & \begin{tabular}[c]{@{}l@{}}Class model derived from UML diagrams -- Boundary (Interface),  \\ Control (business logic) \\ and Entity (mapped to database table).\end{tabular} & Entities separated as microservices. \\ \hline
    59 & Set of Business Processes (BP) to generate the dependency matrix. & Set of clusters derived from dependency matrix. \\ \hline
    69 & \begin{tabular}[c]{@{}l@{}}Data flow diagrams (DFD) as the input. Relationship matrix between \\ primitive functions and data storage for extraction.\end{tabular} & Primitive functions grouping into microservices. \\ \hline
    73 & \begin{tabular}[c]{@{}l@{}}BPEL of the system convert to Subject, Verb, Object table\\  to obtain system vocabulary trees.\end{tabular} & \begin{tabular}[c]{@{}l@{}}System operations grouped as microservices \\ derived from vocabulary trees.\end{tabular} \\ \hline
    74 & \begin{tabular}[c]{@{}l@{}}Use case, requirements and functionalities. From use cases generate\\ operation/relation tables.\end{tabular} & \begin{tabular}[c]{@{}l@{}}Manually identified microservices from the visualization of \\ operation/relation tables.\end{tabular} \\ \hline
    75 & Product backlog's user stories. & \begin{tabular}[c]{@{}l@{}}Decomposed microservices, backlog diagram, \\ and quality matrices.\end{tabular} \\ \hline
    76 & \begin{tabular}[c]{@{}l@{}}Architecture Domain Language (ADL) to identify bounded \\ context from ADL.\end{tabular} & \begin{tabular}[c]{@{}l@{}}Converted and deployable system with one database per \\ microservice.\end{tabular} \\ \hline
    77 & Feature cards and feature table. & \begin{tabular}[c]{@{}l@{}}Feature partitions, identified microservices based on \\ mapping rules.\end{tabular} \\ \hline
    92 & \begin{tabular}[c]{@{}l@{}}Business processes converted to structural and \\ data dependencies relationship matrix.\end{tabular} & Clustered processes as microservices. \\ \hline
    \end{tabular}}
\label{table:inputOutputDDD}
\end{table}

Study~1 is a semi-automated approach that takes System Specification Artifacts (SSA), such as UML and ER diagrams, use cases, security zones, and entities, in JSON-based machine-readable format as input.
Study~7 uses use-case-to-use-case and use-case-to-database-entity relationship matrices to generate a similarity matrix, which serves as a basis for microservice identification.
As input, <component – attribute> data matrix is used, where components can be the use cases of the system, and attributes are its properties.
Study~8 uses business requirements, features, and stored procedure/business logic mapping for features as input.
Then, a microservice discovery table (MDT) is created with system requirements, corresponding features of interest in the source code, and the stored procedures that implement the business logic.
This table is then used as the ground for microservice discovery.
The control, semantic, data, and organizational dependencies between business processes represented in a matrix format with a dependency score matrix are used in Study~21 as input to the microservice extraction.
Studies~40, 41, 48, 54, 75, and 76 use domain artifacts, such as UML, DFD, ADL, BPEL, and use case diagrams as input. 
Studies~40 and 41 produce bounded contexts identified as microservice candidates as outputs.
Studies~58 and 68 follow the same pattern and produce DFD and entity groping, respectively, as output. 
In contrast, the result of Study~75 is a set of matrices and microservices. 
The matrices indicate the quality measures of extracted microservices in terms of complexity, coupling, cohesion, interface count, and estimated development time.
Study~76, as output, provides a converted and deployable system with a repository and database per microservice.
In Study~50, business requirements and functionalities are divided into task levels, and a dependency matrix is created for microservice identification. 
Studies~59, 69, and 92 use matrix-based representations derived from business processes, while DFDs are used to extract the microservices.
A table-based representation of domain artifacts is input to Studies~73, 74, and 77. 
Business Process Execution Language (BPEL) models are used in Study~73 to derive subject-verb-object relationship table.
This table is used as a vocabulary to identify system operations. 
These system operations are used as output microservices. 
In Study~74, use cases are used to construct operation/relations tables for requirements and functionalities.
This table visually represents the identified microservices and is the output of the approach.
Study~77 is grounded in system features.
It has additional input of feature cards, assigning weight to features. 
Microservice candidates are produced as output, resulting from feature table analysis using predefined rules.

{
\scriptsize
\begin{longtable}[t]{|l|l|l|}
    \caption{Inputs and outputs of static analysis approaches}
    \label{table:inputOutputStatic}
    \\ \hline
    \multicolumn{1}{|c|}{\textbf{\begin{tabular}[c]{@{}c@{}}Study\\  ID\end{tabular}}} & \multicolumn{1}{c|}{\textbf{Input and/or intermediate representation}} & \multicolumn{1}{c|}{\textbf{Output}} \\ \hline
    11 & \begin{tabular}[c]{@{}l@{}}Source code. AST and Topic based strength between components \\ as a graph\end{tabular} & \begin{tabular}[c]{@{}l@{}}Components partitions as \\ microservice candidates\end{tabular} \\ \hline
    27 & \begin{tabular}[c]{@{}l@{}}Source code and SQL queries  are represented as Classes to  business \\ object relationship matrix, cosine similarity matrix contains the semantic \\ similarity between classes, subtypes, and reference relationship matrices \\ generated by analysis of the class relationship graph generated by the \\ Mondrian tool\end{tabular} & \begin{tabular}[c]{@{}l@{}}Classes partitioned into \\ clusters as recommanded \\ microservices\end{tabular} \\ \hline
    28 & \begin{tabular}[c]{@{}l@{}}Classes to business object relationship matrix, cosine similarity matrix \\ contains the semantic similarity between methods, method call \\ relationship matrix\end{tabular} & \begin{tabular}[c]{@{}l@{}}Methods partitioned into \\ clusters as recommended \\ microservices\end{tabular} \\ \hline
    29 & \begin{tabular}[c]{@{}l@{}}Source code as primary input. Source code represented as a graph, class \\ as a node and edges as calls between classes.  Additional classes and entry \\ point matrix, classes vs number of common entry points matrix and class \\ inheritance matrix\end{tabular} & Cluster assignment matrix \\ \hline
    31 & \begin{tabular}[c]{@{}l@{}}Source code to MODISCO tool to get the system model as abstract syntax \\ tree\end{tabular} & \begin{tabular}[c]{@{}l@{}}Visual representation of EJB \\ clusters and microservices\end{tabular} \\ \hline
    32 & \begin{tabular}[c]{@{}l@{}}Source code and repository represent as a graph. Graph vertices as \\ classes/interfaces, edges as static and evolutionary coupling\end{tabular} & \begin{tabular}[c]{@{}l@{}}Set of clusters as MS candidates\\ to compare with authoritative \\ decomposition\end{tabular} \\ \hline
    33 & \begin{tabular}[c]{@{}l@{}}Source code \& DB of the application under analysis and a set of  proposed \\ microservices in JSON format, in which each microservice is described by \\ the set of classes that compose that microservice. Source code represents as\\  abstract syntax tree for structural data extraction\end{tabular} & \begin{tabular}[c]{@{}l@{}}Data base and source code \\ refactored as microservices \\ with  communication interfaces\end{tabular} \\ \hline
    44 & \begin{tabular}[c]{@{}l@{}}Source code as set of programs and data (data access write, read \\ operations) represent as a graph\end{tabular} & \begin{tabular}[c]{@{}l@{}}Visualize candidates of \\ microservices as the list of \\ programs and data using \\ city metaphor\end{tabular} \\ \hline
    51 & \begin{tabular}[c]{@{}l@{}}Source code and repository history represented as static, semantic and \\ evolutionary coupling graphs\end{tabular} & \begin{tabular}[c]{@{}l@{}}Classes of clusters as \\ recommended microservices\end{tabular} \\ \hline
    53 & System dependency graph of source code and database & \begin{tabular}[c]{@{}l@{}}Graph communities as \\ recommended microservices\end{tabular} \\ \hline
    55 & Source code is the input. Arcan tool create system dependency graph & \begin{tabular}[c]{@{}l@{}}The semantics with associated \\ Java classes as microservices\end{tabular} \\ \hline
    58 & DB tables and table attributes for topic detection & \begin{tabular}[c]{@{}l@{}}Clustered tables as \\ recommended microservices\end{tabular} \\ \hline
    61 & System functionalities and persistent domain entities. & \begin{tabular}[c]{@{}l@{}}Clustered domain entities as \\ identified microservices\end{tabular} \\ \hline
    66 & Structural similarity and semantic similarity matrix & \begin{tabular}[c]{@{}l@{}}Classes are grouped into \\ microservices and identified \\ outlier classes\end{tabular} \\ \hline
    67 & Classes in the source code. & \begin{tabular}[c]{@{}l@{}}Classes grouped into \\ microservices based on \\ dependencies\end{tabular} \\ \hline
    70 & \begin{tabular}[c]{@{}l@{}}Open API specification based API details to generate API similarity \\ graph\end{tabular} & \begin{tabular}[c]{@{}l@{}}API clusters as recommended \\ microservices\end{tabular} \\ \hline
    79 & \begin{tabular}[c]{@{}l@{}}Source code for reverse engineering to obtain layered architecture meta \\ model for class clustering based on structural and data similarity\end{tabular} & \begin{tabular}[c]{@{}l@{}}Clustered classes as \\ recommended microservices\end{tabular} \\ \hline
    80 & Source code represented as class dependency graph & \begin{tabular}[c]{@{}l@{}}Visualization of graph clusters \\ as candidate microservices\end{tabular} \\ \hline
    85 & \begin{tabular}[c]{@{}l@{}}Open API specification based API details and reference vocabulary \\ details to calculate semantic similarity\end{tabular} & \begin{tabular}[c]{@{}l@{}}API mappings as recommended \\ microservices\end{tabular} \\ \hline
    86 & \begin{tabular}[c]{@{}l@{}}Source code to derive logical, semantic, and contributor coupling \\ graphs\end{tabular} & \begin{tabular}[c]{@{}l@{}}Clustered classes as \\ recommended microservices\end{tabular} \\ \hline
    87 & \begin{tabular}[c]{@{}l@{}}Open API specification based API details to extract operation names \\ for semantic similarity measures\end{tabular} & \begin{tabular}[c]{@{}l@{}}Clustered operation names as \\ recommended microservices\end{tabular} \\ \hline
    89 & \begin{tabular}[c]{@{}l@{}}Source code to analyse static and semantic relationships using machine \\ learning techniques to generate graph-based representation of the system\end{tabular} & \begin{tabular}[c]{@{}l@{}}Clustered classes as \\ recommended microservices\end{tabular} \\ \hline
    90 & \begin{tabular}[c]{@{}l@{}}Source code to extract the methods and code embedding model using \\ neural network model (code2vec) to cluster based on semantic similarity\end{tabular} & \begin{tabular}[c]{@{}l@{}}Clustered classes as \\ recommended microservices\end{tabular} \\ \hline
    91 & \begin{tabular}[c]{@{}l@{}}Source code and database to generate dependency graph of classes, \\ facades, and db tables as vertices and call relationships as edges.\end{tabular} & \begin{tabular}[c]{@{}l@{}}Identified microservice \\ candidates from dependency\\  graph\end{tabular} \\ \hline
\end{longtable}
}

For static analysis approaches, the main inputs are source code, database artifacts, and code repository histories.
These inputs are processed and converted into machine-readable formats for further analysis.
Most studies represent the source code as a graph or matrix-based abstractions, which are then used to discover microservices.
Specifically, graph or matrix-based clustering, genetic, and community detection algorithms are used.
As output, these approaches often provide clusters that define candidate microservices.
\Cref{table:inputOutputStatic} summarizes the details of inputs and outputs used by static analysis approaches.
Again, only the studies with detailed descriptions of the inputs and outputs are included in the table.

Inputs and outputs used by dynamic analysis approaches and constructed intermediate representations of systems are detailed in \Cref{table:inputOutputDynamic}.
These approaches often perform statistical studies over the system’s performance data before identifying its constituent parts or microservices.
System logs are usually collected using instrumented source code.
The latter is also used to conduct use cases and functional testing of the original and reengineered systems.

\begin{table}[t]
\captionof{table}{Inputs and outputs of dynamic analysis approaches}
\noindent\adjustbox{max width=\textwidth}{%
\begin{tabular}{|l|l|l|}
\hline
\textbf{Study ID} & \multicolumn{1}{c|}{\textbf{Input and/or intermediate representation}}                                                                                                          & \textbf{Output}                                                                                                                                                                                                    \\ \hline
2                 & \begin{tabular}[c]{@{}l@{}}Log files collected after AOP based\\ instrumenting feed to DISCO tool to \\ obtain graphical representation of \\ processes.\end{tabular}     & \begin{tabular}[c]{@{}l@{}}Multiple decomposition options with matrix based \\ ranking for solution selection.\end{tabular}                                                                                        \\ \hline
5                 & \begin{tabular}[c]{@{}l@{}}Web server access logs to analyse \\ URI invokes.\end{tabular}                                                                                 & \begin{tabular}[c]{@{}l@{}}URI frequency and mean request \\ response time (MRRT) based clusters.\end{tabular}                                                                                                     \\ \hline
6                 & \begin{tabular}[c]{@{}l@{}}Web server access logs to analyse \\ URI invokes.\end{tabular}                                                                                 & Response size and time based clusters.                                                                                                                                                                             \\ \hline
13                & \begin{tabular}[c]{@{}l@{}}Monitoring logs generated using Kieker \\ with full business operations coverage.\end{tabular}                                                 & \begin{tabular}[c]{@{}l@{}}Method invocation logs with time and \\ frequency as inputs for node attribute network.\end{tabular}                                                                                    \\ \hline
24                & \begin{tabular}[c]{@{}l@{}}System logs with major functionality and \\ use case coverage.\end{tabular}                                                                    & \begin{tabular}[c]{@{}l@{}}Call graph generated using DISCO tool to \\ combine with static analysis results of business \\ objects and operations for clustering.\end{tabular}                                     \\ \hline
25                & \begin{tabular}[c]{@{}l@{}}System logs with major functionality \\ and use case coverage.\end{tabular}                                                                    & \begin{tabular}[c]{@{}l@{}}Call graph generated using DISCO tool to combine \\ with business objects to identify single entry single \\ exit (SESE) to derive frequently execution patterns\\  (FEP).\end{tabular} \\ \hline
26                & \begin{tabular}[c]{@{}l@{}}Execution logs collected by simulating \\ user behavior using Selenium scripts.\end{tabular}                                                   & Call graphs related to executions.                                                                                                                                                                                 \\ \hline
39                & \begin{tabular}[c]{@{}l@{}}Collected execution traces using Kieker \\ tool with pre-defined functional test suite.\end{tabular}                                           & \begin{tabular}[c]{@{}l@{}}Grouped functional atoms after applying NSGA II \\ on identified functional atoms from execution traces.\end{tabular}                                                                   \\ \hline
42                & \begin{tabular}[c]{@{}l@{}}Usecase based runtime logs to identify direct\\ and indirect call relationships to generate \\ similarity matrix between classes.\end{tabular} & Clustered set of class partitions based on similarity.                                                                                                                                                             \\ \hline
43                & \begin{tabular}[c]{@{}l@{}}Usecase based log collected from instrumented \\ source code to generate calling context tree.\end{tabular}                                    & \begin{tabular}[c]{@{}l@{}}Classes as clusters derived after combining  dynamic \\ data with static information.\end{tabular}                                                                                      \\ \hline
46                & \begin{tabular}[c]{@{}l@{}}Live monitoring and visualization using \\ ExploreViz application.\end{tabular}                                                                & \begin{tabular}[c]{@{}l@{}}Identified bounded context by static analysis and \\ ExploreViz visualized resutls.\end{tabular}                                                                                        \\ \hline
52                & \begin{tabular}[c]{@{}l@{}}Operational data (entry points, classes, queries)\\  collected using Silk tool.\end{tabular}                                                   & \begin{tabular}[c]{@{}l@{}}Classes as clusters after combining results with system \\ static data.\end{tabular}                                                                                                    \\ \hline
57                & \begin{tabular}[c]{@{}l@{}}Systems logs to analyze statistics and invoke \\ relationships to generate the call graph with \\ dynamic tracing frequencies.\end{tabular}    & Clustered microservice partitions                                                                                                                                                                                  \\ \hline
72                & \begin{tabular}[c]{@{}l@{}}Collected traces after AOP based instrumenting \\ to feed into DISCO tool.\end{tabular}                                                        & \begin{tabular}[c]{@{}l@{}}Microservices identified after visually inspecting \\ the tool generated call graphs.\end{tabular}                                                                                      \\ \hline
83               & \begin{tabular}[c]{@{}l@{}}Data on frequency of method invokes collected\\ by instrumenting the source code.\end{tabular}                                                 & \begin{tabular}[c]{@{}l@{}}Identified microservice boundaries after combining \\ with static details of the source code.\end{tabular}                                                                              \\ \hline
88               & \begin{tabular}[c]{@{}l@{}}Log files generated after instrumenting the \\ source code and executing test cases.\end{tabular}                                              & Idenfitied microservices after execution traces analysis                                                                                                                                                           \\ \hline
\end{tabular}}
\label{table:inputOutputDynamic}
\end{table}

\subsubsection{RQ2.5) How the reengineered systems are evaluated?}

Once the microservice extraction is completed, it can be evaluated with respect to different aspects.
From the functional viewpoint, all the required features and functions provided by the legacy system should be available in the migrated system. 
Moreover, the system's performance should not be affected due to migration. 
Furthermore, the resulting system must preserve specific quality attributes, such as modularity, loose coupling, high cohesion, and granularity.
In the existing literature, multiple criteria were adopted to evaluate the reengineered systems. 
Next, we provide a detailed discussion of how migrated microservices get evaluated.

According to the literature, multiple techniques were used to evaluate the reengineered systems, including  
manual expert evaluations,
prototype implementations, 
industrial case studies,
cross-system evaluations, and 
property assessments.

The basic approach for validating the refactored system is via expert opinions, which can be carried out directly by evaluating the refactored system by experts or indirectly by comparing the resulting system with expert-extracted solutions.
Prototyping is another approach in which the proposed reengineering technique is applied over one or multiple open-source systems.
In contrast, in industrial case studies, an existing technique gets evaluated over industry applications. 
Cross-system evaluation is a highly used technique in which the proposed solution gets cross-compared with the available state-of-the-art techniques to check if the new solution is superior. 
Property measuring is another widely used technique. 
Properties like modularity, quality of decomposition, and runtime performance of the original and reengineered systems are calculated and compared. 
Moreover, a few studies have considered hyperparameter optimization~\cite{SellamiDBSCAN, AlDebagyPNewDecompositionMethodHyperParameter}, where reengineering technique configurations are evaluated for performance tuning.
	
The properties used to measure the quality of the reengineered systems can broadly be categorized into seven categories: 
runtime performance, 
modularity, 
coupling, 
cohesion, 
independence of functionality and evolvability, 
quality of decomposition, 
and other criteria.
We discuss these categories below.

\paragraph{Runtime performance}
Runtime performance analyzes the parameters of the reengineered system during the execution phase. 
These properties can be directly compared with the monolithic application. 
The specific properties used in such evaluations are listed in \Cref{table:runTimePeformanceIndicators}.

\begin{table}[t]
\captionof{table}{Evaluation of runtime performance}
\noindent\adjustbox{max width=\textwidth}{%
    \begin{tabular}{|l|l|l|}
    \hline
    \multicolumn{1}{|c|}{\textbf{Property}} & \multicolumn{1}{c|}{\textbf{Description}} & \multicolumn{1}{c|}{\textbf{Study IDs}} \\ \hline
    Latency & \begin{tabular}[c]{@{}l@{}}Time to perform user action as the average latency for all the \\ requests by all the users.\end{tabular} & 53 \\ \hline
    Throughput & \begin{tabular}[c]{@{}l@{}}Number of requests completed per second, averaged over the entire \\ runtime of the program.\end{tabular} & 53 \\ \hline
    Scalability & Resource usage over time. & 24, 25, 26 \\ \hline
    Availability & Probability that a system is operational at a given time. & 24, 25, 26 \\ \hline
    Efficiency gain & \begin{tabular}[c]{@{}l@{}}Proportion of total time taken to process all the requests by the legacy \\ system to that by the microservice system.\end{tabular} & 24, 25 \\ \hline
    Network overhead & \begin{tabular}[c]{@{}l@{}}Network congestion due to data volume and exchange frequency after\\ migration.\end{tabular} & 82, 83 \\ \hline
    \end{tabular}}
\label{table:runTimePeformanceIndicators}
\end{table}

\paragraph{Modularity}
Modularity measures how well components of a system can be distinguished, decomposed, and recombined.
Multiple modularity measures were encountered in the literature.
They are listed in \Cref{table:modularityEvaluation}.

\begin{table}[t]
\captionof{table}{Evaluation of modularity}
\noindent\adjustbox{max width=\textwidth}{%
    \begin{tabular}{|l|l|l|}
    \hline
    \multicolumn{1}{|c|}{\textbf{Property}} & \multicolumn{1}{c|}{\textbf{Description}} & \multicolumn{1}{c|}{\textbf{Study IDs}} \\ \hline
    Modularity & Evaluate the quality of clusters in a graph. & 29, 51, 80 \\ \hline
    Structural modularity & \begin{tabular}[c]{@{}l@{}}Structural soundness of the cluster. Evaluate the \\ modularity from structural view point.\end{tabular} & \begin{tabular}[c]{@{}l@{}}7, 11, 29, 39, 42,\\  43, 66, 89\end{tabular} \\ \hline
    Conceptual modularity & \begin{tabular}[c]{@{}l@{}}Evaluate the modularity from conceptual view \\ point.\end{tabular} & 11, 39, 89 \\ \hline
    Mean cluster factor & \begin{tabular}[c]{@{}l@{}}Quality of the partitions by analyzing their \\ interconnectivity and intraconnectivity.\end{tabular} & 51 \\ \hline
    Feature modularization & \begin{tabular}[c]{@{}l@{}}Criterion to optimize the responsibility of microservices. \\ Predominant feature number is the number of occurrences \\ of the most common feature divided by the sum of all feature\\ occurrences. Feature modularization is the sum of the \\ predominant feature number in every microservice divided \\ by the number of distinct features.\end{tabular} & 82, 83 \\ \hline
    \end{tabular}}
\label{table:modularityEvaluation}
\end{table}

In most of the studies, modularity calculation has been conducted based on the method by~\citet{NewmanGModularity}. 
Feature modularity is a measure of feature distribution across the system derived from the notion of the single responsibility per microservice.

\paragraph{Coupling}
Loose coupling, or the degree of interdependence between microservices, is one of the main attributes studied in microservice systems. 
Multiple measures related to coupling were observed in the literature.
They are listed in \cref{table:couplingEvaluation}.
Combined measures of absolute importance, absolute dependence, and microservice interdependence were also used in the evaluations.

\begin{table}[t]
\captionof{table}{Evaluation of coupling}
\noindent\adjustbox{max width=\textwidth}{%
    \begin{tabular}{|l|l|l|}
    \hline
    \multicolumn{1}{|c|}{\textbf{Property}} & \multicolumn{1}{c|}{\textbf{Description}} & \multicolumn{1}{c|}{\textbf{Study IDs}} \\ \hline
    Coupling (between microservices) & Measures the level of interaction between services. & \begin{tabular}[c]{@{}l@{}}2, 49, 53, 61, \\ 72, 82, 83\end{tabular} \\ \hline
    Structural coupling & \begin{tabular}[c]{@{}l@{}}Structural relationship between classes in different\\ modules.\end{tabular} & 24, 25, 27, 28 \\ \hline
    Afferent coupling & \begin{tabular}[c]{@{}l@{}}Service responsibility on other services. Measure of the \\ number of classes in other services depending on classes \\ within a service.\end{tabular} & 48, 77 \\ \hline
    Efferent coupling & \begin{tabular}[c]{@{}l@{}}Indicates the number of outgoing dependencies. \\ The number of classes in a service depends on \\ the classes in other services.\end{tabular} & 48, 77 \\ \hline
    Internal coupling & \begin{tabular}[c]{@{}l@{}}Degree of direct and indirect dependencies between \\ classes within a microservice.\end{tabular} & 67 \\ \hline
    External coupling & \begin{tabular}[c]{@{}l@{}}Degree of direct and indirect dependencies in a class \\ that belong to a candidate microservice to external classes.\end{tabular} & 67 \\ \hline
    Coupling at method level & \begin{tabular}[c]{@{}l@{}}Measures the inter-connectivity between system \\ components. Used in method-based modeling.\end{tabular} & 13 \\ \hline
    \begin{tabular}[c]{@{}l@{}}Absolute importance \\ of a service (AIS)\end{tabular} & \begin{tabular}[c]{@{}l@{}}Number of clients that invoked at least one operation of \\ the microservice interface.\end{tabular} & 75 \\ \hline
    \begin{tabular}[c]{@{}l@{}}Absoulte dependence \\ of a service (ADS)\end{tabular} & \begin{tabular}[c]{@{}l@{}}Number of microservices that invoked at least one \\ operation of the service.\end{tabular} & 75 \\ \hline
    Microservice interdependence & Number of interdependent service pairs. & 75 \\ \hline
    Instability index & \begin{tabular}[c]{@{}l@{}}The measure of system resilience to change. Calculate as a\\ fraction of afferent coupling and efferent coupling.\end{tabular} & 7, 48, 77 \\ \hline
    \end{tabular}}
\label{table:couplingEvaluation}
\end{table}

\paragraph{Cohesion} 
A basic design principle of a microservice is high cohesiveness.
Classes used to implement a microservice should be highly co-related. 
Highly cohesive microservice governs a single responsibility. 
Interfaces exposed by the microservice and messages used to communicate with the microservice should be well-defined and focus on that single responsibility.
\Cref{table:cohesionEvaluation} explains the cohesion measures used in the literature.

{
\small
\begin{table}[t]
\captionof{table}{Evaluation of cohesion}
\noindent\adjustbox{max width=\textwidth}{%
    \begin{tabular}{|l|l|l|}
    \hline
    \multicolumn{1}{|c|}{\textbf{Property}} & \multicolumn{1}{c|}{\textbf{Description}} & \multicolumn{1}{c|}{\textbf{Study IDs}} \\ \hline
    Cohesion & \begin{tabular}[c]{@{}l@{}}Measure of degree of interconnectedness within a \\ service. Represents the number of static calls\\ within a service over all the static calls.\end{tabular} & \begin{tabular}[c]{@{}l@{}}49, 53, 61, \\ 82, 83\end{tabular} \\ \hline
    Relation cohesion & \begin{tabular}[c]{@{}l@{}}Number of internal relations including \\ inheritance, method invocations, access to\\ class attributes and access via references.\end{tabular} & 7, 48, 77 \\ \hline
    \begin{tabular}[c]{@{}l@{}}Cohesion at message\\ level (CHM)\end{tabular} & \begin{tabular}[c]{@{}l@{}}Cohesiveness at interface messages within\\ a service.\end{tabular} & \begin{tabular}[c]{@{}l@{}}11, 39, 88, \\ 89, 90\end{tabular} \\ \hline
    \begin{tabular}[c]{@{}l@{}}Cohesion at domain\\ level (CHD)\end{tabular} & \begin{tabular}[c]{@{}l@{}}Cohesiveness at domain level of the service.\\ Measures the similarity of functions.\end{tabular} & \begin{tabular}[c]{@{}l@{}}11, 39, 88, \\ 89, 90\end{tabular} \\ \hline
    \begin{tabular}[c]{@{}l@{}}Cohesion at method\\ level (M\_CH)/ Internal \\ cohesion\end{tabular} & \begin{tabular}[c]{@{}l@{}}Method level connectivity in a service. Used \\ for method-based modeling.  Degree of the \\ number of methods invoked against all the methods \\ in a service.\end{tabular} & 13, 67 \\ \hline
    Lack of cohesion & \begin{tabular}[c]{@{}l@{}}Measure of the number of pairs of services \\ not having any dependency between them.\end{tabular} & \begin{tabular}[c]{@{}l@{}}24, 25, 27, \\ 28, 75\end{tabular} \\ \hline
    Density & Internal co-relation degree of each microservice. & 13 \\ \hline
    \end{tabular}}
\label{table:cohesionEvaluation}
\end{table}
}

\paragraph{Independence of functionality and evolvability}
A microservice should be independent, which supports flexible changes in the system that do not affect other services.
Therefore, functional independence is an essential characteristic of microservices. 
Evaluation of functional independence of microservices is done in several ways in the literature, see listed in \Cref{table:functionalEvaluation}.

\begin{table}[t]
\captionof{table}{Evaluation of independence of functionality and evolvability}
\noindent\adjustbox{max width=\textwidth}{%
    \begin{tabular}{|l|l|l|}
    \hline
    \multicolumn{1}{|c|}{\textbf{Property}} & \multicolumn{1}{c|}{\textbf{Description}} & \multicolumn{1}{c|}{\textbf{Study IDs}} \\ \hline
    Interface number & \begin{tabular}[c]{@{}l@{}}Average number of interfaces published\\ by a microservice to other services.\end{tabular} & \begin{tabular}[c]{@{}l@{}}7, 11, 29, 39, 42, \\ 66, 88, 89\end{tabular} \\ \hline
    \begin{tabular}[c]{@{}l@{}}Interaction number\\ Inter partition/internal\\ calls as a percentage/number.\end{tabular} & \begin{tabular}[c]{@{}l@{}}Percentage of calls between two microservices/ \\ the number of calls between two services.\end{tabular} & \begin{tabular}[c]{@{}l@{}}7, 11, 42, 43, 53, \\ 88\end{tabular} \\ \hline
    OPN - operation number & \begin{tabular}[c]{@{}l@{}}Number of operations provided by extracted \\ microservices.\end{tabular} & 88 \\ \hline
    \begin{tabular}[c]{@{}l@{}}Internal co-change\\ frequency\end{tabular} & \begin{tabular}[c]{@{}l@{}}Frequency of entity changes \\ inside a service based on revision history.\end{tabular} & 39 \\ \hline
    \begin{tabular}[c]{@{}l@{}}External co-change \\ frequency\end{tabular} & \begin{tabular}[c]{@{}l@{}}Frequency of entity change in \\ different services based on revision\\ history.\end{tabular} & 39 \\ \hline
    \begin{tabular}[c]{@{}l@{}}Number of duplicate\\ classes\end{tabular} & \begin{tabular}[c]{@{}l@{}}Used to identify the appropriate decomposition \\ from multiple options. Number of classes duplicated \\ in different services.\end{tabular} & 2, 72 \\ \hline
    \begin{tabular}[c]{@{}l@{}}Frequency of external \\ calls\end{tabular} & \begin{tabular}[c]{@{}l@{}}Low frequency of external calls optimize the \\ performance and reduce delays. Fraction of number of \\ calls over the number of classes in a microservice.\end{tabular} & 72 \\ \hline
    \end{tabular}}
\label{table:functionalEvaluation}
\end{table}

In addition, the fraction between external change frequency (across services) to internal change frequency (within services) is known as the REI ratio.
Ideally, changes inside a service should be higher than those across the services. 
Therefore, the value is expected to be less than one. 
Smaller values indicate the services tend to evolve independently~\cite{JinExecutionTracesFosci} (Study 48).

\paragraph{Quality of decomposition}
The measures from this category assess the quality of the functional distribution across the microservices.
This distribution can be, for instance, in terms of use cases, operations, or classes. 
The existing measures are listed in \Cref{table:qualityOfDecomposition}.

\begin{table}[t]
\captionof{table}{Evaluation of decomposition quality}
\noindent\adjustbox{max width=\textwidth}{%
    \begin{tabular}{|l|l|l|}
    \hline
    \multicolumn{1}{|c|}{\textbf{Property}} & \multicolumn{1}{c|}{\textbf{Description}} & \multicolumn{1}{c|}{\textbf{Study IDs}} \\ \hline
    Business context purity & \begin{tabular}[c]{@{}l@{}}Indicates business use case distribution \\ across the services. Mean entropy of \\ business use cases per partition.\end{tabular} & 7, 42, 43, 53 \\ \hline
    DB transactional purity & \begin{tabular}[c]{@{}l@{}}Measure of distributed transactions. Per each DB table,\\ calculate the partitions access the table to get the entropy.\end{tabular} & 53 \\ \hline
    \begin{tabular}[c]{@{}l@{}}Non-extreme distribution/ \\ classes per microservice\end{tabular} & \begin{tabular}[c]{@{}l@{}}Measure how evenly distributed the classes among the \\ microservices.\end{tabular} & \begin{tabular}[c]{@{}l@{}}2, 7, 29, 42, \\ 66, 72\end{tabular} \\ \hline
    \begin{tabular}[c]{@{}l@{}}Number of classes per\\ microservice\end{tabular} & Used to identify the better partition recommendations. & 2, 72 \\ \hline
    \begin{tabular}[c]{@{}l@{}}Precision, Recall and \\ F measure\end{tabular} & \begin{tabular}[c]{@{}l@{}}Measurements of reasonable decomposition over actual \\ decomposition.\end{tabular} & \begin{tabular}[c]{@{}l@{}}70, 80,  85, \\ 87, 89, 90\end{tabular} \\ \hline
    Code redundancy rate & \begin{tabular}[c]{@{}l@{}}Based on more code after migration indicates high\\ communication concept. Code volume difference in original \\ and migrated system over original application code volume,\end{tabular} & 49 \\ \hline
    Domain redundancy rate & \begin{tabular}[c]{@{}l@{}}Based on better decomposition avoids duplication of \\ responsibilities.\end{tabular} & 86 \\ \hline
    Team size reduction & \begin{tabular}[c]{@{}l@{}}Microservice improves the organization of development team.\\ Team size of each service is the number of functions provided\\ by the partition.\end{tabular} & 49, 86 \\ \hline
    Reuse & \begin{tabular}[c]{@{}l@{}}Reuse is measured by the relationship between identified \\ services and the legacy system users e.g., API calls, UI \\ interactions, etc. Analysis needs to be conducted in migrated system \\ to calculate this property.\end{tabular} & 82 \\ \hline
    \end{tabular}}
\label{table:qualityOfDecomposition}
\end{table}

One of the central properties for assessing the quality of a microservice decomposition is transactional database purity.
This measure prioritizes decompositions with dedicated databases per microservice. 
The notion of entropy is often used to assess transactional purity.
The transactional purity of a system is then established in terms of the purities across all the database tables, where smaller entropy values indicate high transactional purity~\cite{NitinARKCargo}.

\paragraph{Other measures}
MoJoSim and MoJoFM are used to evaluate a microservice-based architecture against a reference architecture (e.g., against an expert-identified architecture).
It is calculated by measuring the minimum number of operations (e.g., move or join) required to transform the identified microservice architecture to the reference architecture~\cite{EskiBAnAutomaticExtractionApproach, ZaragozaSAADLeveragingLayeredArchit}. 
API division accuracy~\cite{Sun2022ExpertSF} is a measure to calculate the efficiency of API identification. 
It calculates the accuracy by relating the correctly identified API against all APIs. 
The cluster-to-cluster coverage (c2ccvg) measures the degree of overlap of the implementation-level entities between two clusters~\cite{ZaragozaSAADLeveragingLayeredArchit}.

In addition, existing studies perform hyperparameter optimization to optimize the microservice decomposition.
They explore multiple alternative decompositions to identify optimal ones with respect to the properties discussed above~\cite{SellamiDBSCAN,AlDebagyPNewDecompositionMethodHyperParameter}.  
Silhouette coefficient (SC) is used to evaluate the performance of the clustering algorithms~\cite{AlDebagyPNewDecompositionMethodHyperParameter}.

Existing applications have been used to implement and evaluate the proposed reengineering solutions. 
Most of the reengineered systems were Java-based, with limited PHP systems identified. 
Applications reengineered in at least two works are listed in \Cref{table:evaluatedApplications}.

{
\small
\begin{table}[t]
\captionof{table}{Evaluated applications}
\noindent\adjustbox{max width=\textwidth}{%
    \begin{tabular}{|l|l|l|}
    \hline
    \multicolumn{1}{|c|}{\textbf{Application}} & \multicolumn{1}{c|}{\textbf{Study IDs}} & \multicolumn{1}{c|}{\textbf{Technology}} \\ \hline
    JPetStore & 7, 11, 13, 39, 42, 53, 57, 66, 80, 88, 90 & Java \\ \hline
    Acme Air & 6, 7, 29, 53, 66, 80 & Java \\ \hline
    Cargo Tracking System & 1, 7, 48, 49, 77, 80 & Java \\ \hline
    Daytrader & 29, 43, 53, 55, 57, 66 & Java \\ \hline
    Springblog & 39, 80, 88, 90 & Java \\ \hline
    Jforum & 39, 88, 89, 90 & Java \\ \hline
    Apache Roller & 39, 88, 90 & Java \\ \hline
    Spring boot pet clinic & 44, 66, 89 & Java \\ \hline
    E-commerce system & 49, 58 & Java \\ \hline
    Microservices event sourcing & 66, 70 & Java \\ \hline
    Kanban board & 66, 70 & Java \\ \hline
    \begin{tabular}[c]{@{}l@{}}TFWA (Teachers Feedback \\ Web Application)\end{tabular} & 5, 7 & Java \\ \hline
    Train Ticket Microservice Benchmark & 12, 88 & Java \\ \hline
    Plants by WebSphere & 29, 53 & Java \\ \hline
    Sugar CRM & 24, 25, 26, 27 & PHP \\ \hline
    Church CRM & 24, 25, 26, 27 & PHP \\ \hline
    \end{tabular}}
\label{table:evaluatedApplications}
\end{table}
}

Exhaustive evaluations have been conducted in a few studies, where the presented solution has been compared against the existing migration frameworks. 
In the early stages of migration, solutions were compared against the service identification-based systems instead of microservice extraction-based systems. 
Service identification-based systems are not included in \Cref{table:crossComparedSystems}, which lists the details of cross-compared evaluations.

{
\small
\begin{table}[t]
\captionof{table}{Cross-system evaluation frameworks}
\noindent\adjustbox{max width=\textwidth}{%
    \begin{tabular}{|l|l|l|}
    \hline
    \textbf{Study ID} & \textbf{Short Name} & \textbf{Cross-compared studies} \\ \hline
    1 & Service Cutter & 7, 48, 52, 70, 75, 77 \\ \hline
    29 & CoGCN & 7, 42, 53 \\ \hline
    39 & FoSCI & 7, 13, 42, 53 \\ \hline
    42 & Mono2Micro & 7, 53 \\ \hline
    48 & DFD & 7, 13, 77 \\ \hline
    85 & API analysis & 7, 48, 77 \\ \hline
    86 & MEM & 7, 39, 53, 88 \\ \hline
    \end{tabular}}
\label{table:crossComparedSystems}
\end{table}
}

\subsection{RQ3) What are the challenges and limitations of the existing microservice migration methods?}

There is no standard method for migrating or evaluating systems.
Therefore, depending on the underline system and approaches followed, different outcomes and results are achieved, which are often hard to compare.
Multiple challenges associated with migrating, developing, deploying, and maintaining the migrated systems have been identified.
First, deciding to embarque into the legacy migration project is an organizational challenge due to the facts listed below.

\begin{itemize}
\item Business owner approvals; resistance from business owners to change a stable system 
\item Software development process; resistance to change the software development process, for instance, from the Waterfall to the Agile process
\item Team restructuring and collaboration; unwillingness to undergo team-level restructuring to develop and maintain the new system
\item Cost of migration; cost incurred from acquiring resources (e.g., human resources), design, and development to establish and maintain the infrastructure of the reengineered system
\end{itemize}
    
\noindent
Moreover, various technical challenges have been identified, as summarized below.

\begin{itemize}
\item 
Technical limitations, for example, lack of expert knowledge and tools for migration. DevOps and cloud adoption are required to obtain the maximum benefits of microservices, demanding additional technical expert knowledge.
\item 
Identification of the optimal set of microservices with the appropriate granularity. The success of a microservice reengineering project depends on a suitable tradeoff between the granularity and the magnitude of the microservices. There is no standard approach to identifying these factors. However, an optimal number of services is required for scalability and long-term maintainability.  
\item 
Decomposition of strongly coupled APIs, classes, and interfaces. Management of shared classes increases the maintainability costs.
\item 
Quality. There is no standard way to measure the quality of a decomposed microservice system.
\item 
Management of the communication between the services. The additional communication cost between the migrated system's components affects its performance.
\item Infrastructure for maintenance, monitoring, and security. Distributed deployment of a microservice system requires additional security and infrastructure supportability.
\item Data consistency. Decomposition of the database layer can lead to data inconsistencies between the services as opposed to the existing centralized persistence layer.
\item Statefulness. A stateful system produces outputs that depend on the state generated in previous interactions~\cite{AndreiCOWAMigratingToMS}. Unlike monolithic systems, managing statefulness in a microservice-oriented system is challenging due to its distributed nature.
\end{itemize}

\noindent
Multiple limitations restrict existing migration frameworks. 
The primary limitation is the unavailability of a standard mechanism for optimal migration and assessing the decomposition quality. 
Moreover, based on the approach used, common limitations identified across the service migration systems are listed in \Cref{table:limitationsList}.

\begin{table}[t]
\captionof{table}{Limitations of existing service migration systems}
\noindent\adjustbox{max width=\textwidth}{%
    \begin{tabular}{|l|l|l|}
    \hline
    \multicolumn{1}{|c|}{\textbf{Limitation}} & \multicolumn{1}{c|}{\textbf{Description}} & \multicolumn{1}{c|}{\textbf{Study IDs}} \\ \hline
    \begin{tabular}[c]{@{}l@{}}A significant effort to transform \\ software system artifacts\end{tabular} & \begin{tabular}[c]{@{}l@{}}System artifacts need to convert to JSON based format\\ for further processing.\end{tabular} & 1 \\ \hline
    Limitations in existing tools & \begin{tabular}[c]{@{}l@{}}Tools incorporated with the process, e.g., Disco,\\ identify processes incorrectly that directly impacts \\ the output.\end{tabular} & 2 \\ \hline
    \multirow{4}{*}{Limitation in system applicability} & Web apps only & 5, 6 \\ \cline{2-3} 
     & Java based systems only & 11, 33 \\ \cline{2-3} 
     & EJB based apps only & 31 \\ \cline{2-3} 
     & Greenfield development only & 7 \\ \hline
    \begin{tabular}[c]{@{}l@{}}Limitations in the approaches and \\ techniques\end{tabular} & \begin{tabular}[c]{@{}l@{}}Gaps in the structural and behavioral analysis and \\ class similarity analysis.\end{tabular} & \begin{tabular}[c]{@{}l@{}}24, 25, \\ 27, 28\end{tabular} \\ \hline
    \multirow{3}{*}{Quality of the artifacts} & \begin{tabular}[c]{@{}l@{}}Proper naming convention should follow in the\\ code base.\end{tabular} & 11 \\ \cline{2-3} 
     & Depends on quality of data flow diagrams(DFD) & 48 \\ \cline{2-3} 
     & Quality of the object oriented source code & 67 \\ \hline
    \multirow{2}{*}{Limitations in database decomposition} & Limited in database interactions and transaction patterns. & 43, 44 \\ \cline{2-3} 
     & Supporting relational databases only. & 76 \\ \hline
    Time for analysis & Accuracy depending on the period of data collection. & 48, 52 \\ \hline
    \end{tabular}}
\label{table:limitationsList}
\end{table}

\section{Discussion and Future Directions}
\label{sec:discussion}

The state-of-art technique for microservice discovery is static analysis, which has been thoroughly explored. 64\% of the studies were based on static analysis.
In particular, structural analysis is the prominent technique over semantic analysis and evolutionary coupling analysis.
CoGCN~\cite{Desai2021GraphNN}, Cargo-AI Guided Dependency Analysis~\cite{NitinARKCargo}, and dependency-based microservice decomposition~\cite{Omar} are the lending studies with structural analysis.
Cargo-AI Guided Dependency Analysis is the prominent structural analysis technique due to the evaluations conducted against benchmark studies. 
Moreover, combined static analysis techniques have shown promising results. 
Microservice identification through topic modeling~\cite{BritoTopicModeling} uses an abstract syntax tree and topic-based strength between components as a graph followed by a modularity-optimized Louvain algorithm, and DBSCAN~\cite{SellamiDBSCAN} uses an abstract syntax tree and semantic similarity as a matrix with a variation of the DBSCAN algorithm.
These are the most notable studies based on structural and semantic analysis.
MEM~\cite{MazlamiExtractionOfMicroservices} is a key study based on evolutionary coupling.
It constructs logical, semantic, and evaluation coupling graphs and uses a minimum spanning tree-based microservice detection algorithm.
Additionally, the automatic extraction approach~\cite{EskiB2018AutomaticExtractionApproach} applied fast community graph clustering to the graph generated using structural and semantic information, and Steinmetz~\cite{JakobOSteinmetz} used a combination of static, semantic, and evolutionary coupling graphs and experimented with multiple clustering algorithms.
These approaches were based on combined static analysis approaches and produced reasonable results.
Moreover, Steinmetz concluded that the Louvain clustering algorithm provides the best results.
In general, static analysis techniques rely highly on existing tools.
However, imprecise program analysis is a major concern for this technique~\cite{NitinARKCargo}.
 
One database per microservice pattern is the preferred approach in the literature~\cite{TyszberwiczIdentifyMSFunctionDecomposition,NitinARKCargo}.
The persistent layer in the source code has been considered while extracting the services~\cite{RefactoringJavaMonolithsFreitas2021}.
However, SQL queries to object mapping have also been considered~\cite{Alwis, Alwis2018FunctionalSplitting,Alwis2019AvailabilityScalability}.
Only one study focuses on identifying microservice candidates from business rules captured in stored procedures~\cite{Barbosa}.
An Object Relational Mapping (ORM) based system has been proposed that relies on specific properties for evaluating the reengineered system~\cite{SantisSilva2022ORM}.
 
Among the two studies on API analysis, MS decomposer~\cite{SunSSHY2022APISimilarity} and API analysis~\cite{BaresiGD2017InterfaceAnalysis}, MS decomposer identifies microservices by creating a topic and response similarity graph followed by K-means clustering.
It evaluates the efficiency of the identified microservices against those discovered by Service Cutter and calculates their properties.

The artifact-driven approaches constitute 23\% of the reviewed studies. 
Service Cutter~\cite{GyselServiceCutter} is one of the pilot studies conducted in microservice identification. 
Hence, it has been used for evaluation in multiple studies. 
The dataflow-driven technique~\cite{ShanshanHZZCJQJZDataflowDrivenApproach} is another prominent artifact-driven approach comprising quality attribute evaluation.
Greenmicro~\cite{BajajGreenMicro}, Microservice Backlog~\cite{FredyEHMicroserviceBacklog}, and the Feature Table approach~\cite{WeiFeatureTable} have shown promising experimental results in comparisons with other migration studies. 
Greenmicro is the latest study in the ADA approach with an extensive cross-system analysis.

Limited experiments have been conducted with dynamic analysis techniques. 
One approach supplies software logs as input to a process mining tool DISCO for further analysis.
However, certain processes have been incorrectly identified by this approach~\cite{TaibiProcessMining, Taibi2020DecompositionAndMetricBasedEvaluation}.
FoSCI~\cite{JinExecutionTracesFosci}, FoME~\cite{Jin2018FoME}, and mono2micro~\cite{KaliaMonoToMicro} are the prominent studies in dynamic analysis.
Moreover, mono2micro is a commercially available product. 
It collects software log traces by executing use cases and identifies unique traces to derive direct and indirect calls to generate a similarity matrix followed by hierarchical clustering.
Furthermore, its strategy has been compared with FoSCI~\cite{JinExecutionTracesFosci}, CoGCN~\cite{Desai2021GraphNN}, Bunch~\cite{AutomaticModularizationOfSoftwareSystemsBunchMitchell2006}, and MEM~\cite{MazlamiExtractionOfMicroservices} to validate the results.
FoSCI uses reduced execution traces to identify functional atoms using the NSGA II multi-objective optimization algorithm.
FoME collects logs from test executions and generates descriptive log traces for clustering and shared class processing.
Both FoSCI and FoME use the Kieker runtime monitoring tool for property evaluation and comparing results against MEM~\cite{MazlamiExtractionOfMicroservices}, LIMBO~\cite{LIMBOInformationTheoreticSoftwareClusteringAndritsos2005}, and WCA~\cite{WCAChatterjee2002}.

Hybrid techniques find their attention in the literature.
Microservice extraction based on knowledge graphs~\cite{ZhidingCJY2022KnowledgeGraph} is a notable study that combines static and artifact-driven analysis. 
It uses source code, database schema, design and API documents to derive a graph with data, module, function, and resource details. 
It then applies the Louvain community detection algorithm to identify microservice candidates.
Multiple hybrid studies of static and dynamic analysis exist~\cite{CaoZ2022NodeAttributedNetwork,MatiasMSBoundaries,EasyAPM,WesleyCLAPM2022ManyObjectiveOptimization,WesleyCLAPM2021MultiCriteriaStrategy}. 
Method invocations were analyzed statistically using call graphs and dynamically using the Kieker run time analysis tool to generate a graph structure for community detection using the Leiden community detection algorithm~\cite{CaoZ2022NodeAttributedNetwork}. 
This is the most comprehensively evaluated hybrid approach cross-compared with FoSCI~\cite{JinExecutionTracesFosci}, the dataflow-driven approach~\cite{ShanshanHZZCJQJZDataflowDrivenApproach}, and the distributed representation approach~\cite{AlDebagy2021AMD}.
MonoBreaker~\cite{MatiasMSBoundaries} analyses the structure of a software project to generate a graph model. 
The graph model is populated with operational data from runtime monitoring followed by the Girvan-Newman algorithm for clustering. 
To analyze the performance, this has been compared against Service Cutter~\cite{GyselServiceCutter} to identify that combining static and dynamic analysis provides better results than static analysis.
Migrating web applications~\cite{EasyAPM} creates a dependency graph using static analysis and enhances it via dynamic analysis.
Microservice candidates are then identified using the K-means clustering algorithm. 
However, this study only evaluates the properties of the reengineered system. 
Other prominent hybrid approaches~\cite{WesleyCLAPM2021MultiCriteriaStrategy,WesleyCLAPM2022ManyObjectiveOptimization} employ multi-objective optimization algorithm NSGA III and perform property evaluations.

Reverse engineering of software systems to derive microservices is rarely used. 
Only two reviewed studies are grounded in reverse engineering of monolithic systems~\cite{EscobarCAREKCCReverseEngineering,ZaragozaSAADLeveragingLayeredArchit}.
We believe that applying reverse engineering techniques to reveal the architecture of a system can support further enhancements of microservice discovery techniques.

Microminer~\cite{AbdellatifTypeBasedApproachML} and the distributed representation of the source code~\cite{AlDebagy2021AMD} have introduced machine learning to microservice extraction. 
Microminer uses a machine learning-based word2vec model with the Louvain community detection algorithm, while the distributed representation of the source code uses a code2vec model with the affinity propagation algorithm.
However, these approaches have no cross-comparison with prominent migration techniques. 
Instead, property calculations have been done to evaluate the proposed solutions.

The evaluation of migrated systems has been mainly based on property calculation. 
Several prominent studies have cross-compared with previous studies~\cite{BajajGreenMicro, FredyEHMicroserviceBacklog, WeiFeatureTable, NitinARKCargo, SellamiDBSCAN, JinExecutionTracesFosci, KaliaMonoToMicro, Jin2018FoME, CaoZ2022NodeAttributedNetwork}.
Service Cutter~\cite{GyselServiceCutter} is the classical migration study used for cross-comparison.
Interface numbers, inter-partition call percentages, and structural modularity are the widely used properties.
Even though coupling has been evaluated in many studies, there is no convergence in the evaluated definitions of this concept. 
Afferent coupling (measuring incoming dependencies) and efferent coupling (outgoing dependencies) are frequently used coupling measurements.
Precision, recall, and F-measure are used for evaluation when a standard decomposition is available for comparison. 
This can be an available microservice system or an expert decomposition result. 
Existing monolithic and microservice-based systems like Kanban\footnote{https://github.com/eventuate-examples/es-kanban-board}, Money Transfer\footnote{https://github.com/cer/event-sourcing-examples}, Piggy Metrics\footnote{https://github.com/sqshq/PiggyMetrics}, Microservices Event Sourcing (MES)\footnote{https://github.com/chaokunyang/microservices-event-sourcing}, Sock Shop\footnote{https://github.com/microservices-demo/microservices-demo} have been used for evaluation~\cite{SunSSHY2022APISimilarity}. 
Limited studies focused on hyperparameter optimization~\cite{SellamiDBSCAN, AlDebagyPNewDecompositionMethodHyperParameter}. 
\citet{YedidaKKMTJLessonsLearnedFromHPO} discussed performance improvements by optimizing hyper-parameters.

The majority of the migration frameworks applied their concepts to monolithic open-source projects.
JPetStore is the most frequently used project for implementation and testing. 
Moreover, Acme Air, Cargo Tracking System, and Daytrader applications were used frequently in the reengineering projects. 
Web-based applications like online shopping systems, learning management systems, banking systems, ERP systems, real-estate applications, web-based IDEs, taxation office systems, and police department systems were also used as proofs of concept. 
Above 80\% of the reengineered in the literature applications are Java-based projects. Database-oriented applications, like stored procedure decompositions, have been discussed in relatively few studies.

Incremental and iterative transitions are the industry-preferred approach for migrating legacy systems to microservices~\cite{BozanLR2020HowToTransitionIncrementally,MichaelHLHMigrationJourneyTowardsMicroservices2021,LevcovitzTVTechniqueForExtractingMicroservices,CarrascpBDMigratingTowardsMicroservices}. 
The behavioral aspects of reengineered systems have not been studied. 
The focus of studies to date was restricted to identifying microservice candidates. 
The runtime performance of the reengineered systems was measured only from the latency, throughput, availability, and network overhead perspectives. 
However, the behavioral aspects, like the coverage of the use cases, have not been considered. 
Moreover, measuring the performance in distributed transactions and validating the data consistency across the microservices has not been evaluated. 
Furthermore, optimal resource utilization in microservices during the execution phase has not been examined. 
Evaluation of dynamic rearrangement of the identified microservices under different system loads is required to increase the efficiency of the migrated systems.

In the literature, identifying microservice candidates from monolithic applications attracts increasing attention. 
However, auto-generating the communication interfaces for microservices has not been considered.
The cost of decomposition in terms of the effort to redesign a functionality has been considered previously~\cite{SantosRComplexityMetricForMicroserviceArchitectureMigration2020}. 
However, a systematic approach has not been proposed to calculate the cost and complexity of the migration process.

\section{Conclusion}
\label{sec:conclusion}

A broad analysis of existing microservice migration research has been considered in this literature review. Initially, 2,238 papers were selected from five research paper libraries. After multiple stages of filtering, 92 primary studies were selected for further analysis. 

The identified studies were analyzed based on multiple perspectives of existing approaches, techniques, tools, data usage, evaluation, limitations, and challenges. We have identified well-explored, state-of-the-art techniques like static analysis and areas with limited focus to date, like dynamic analysis. In addition, the unavailability of convergence in the studies proves that microservice migration research is still in its infancy.

Finally, microservice reengineering is a significant study area that can be improved further. Future studies must focus on exploring new techniques and evaluation strategies for microservice discovery, implementation, deployment, and assessment.

\smallskip
\noindent
\textbf{Acknowledgements.}
This work was in part supported by the Australian Research Council project DP220101516.

\bibliographystyle{plainnat}
\bibliography{LiteratureReview} \label{References}

\appendix

\section{Existing Literature Review Studies}
\label{appendix:litReviewStudies}
	
{
\scriptsize	
		
  		\begin{longtable}{ |c|p{0.91\textwidth}| } 

			\hline
			\textbf{ID} & \textbf{Full reference} \\
			\hline
			LR1 & Schmidt, R., \& Thiry, M. (2020). Microservices identification strategies : A review focused on Model-Driven Engineering and Domain Driven Design approaches. 2020 15th Iberian Conference on Information Systems and Technologies (CISTI), 1-6. \\
			\hline
			LR2 & Schröer, C., Kruse, F., Marx Gómez, J. (2020). A Qualitative Literature Review on Microservices Identification Approaches. In: Dustdar, S. (eds) Service-Oriented Computing. SummerSOC 2020. Communications in Computer and Information Science, vol 1310. Springer, Cham. https://doi.org/10.1007/978-3-030-64846-6\_9 \\
			\hline
			LR3 & M. -D. Cojocaru, A. Uta and A. -M. Oprescu, ``Attributes Assessing the Quality of Microservices Automatically Decomposed from Monolithic Applications,'' 2019 18th International Symposium on Parallel and Distributed Computing (ISPDC), Amsterdam, Netherlands, 2019, pp. 84-93, doi: 10.1109/ISPDC.2019.00021. \\
			\hline
			LR4 & J. Kazanavičius and D. Mažeika, ``Migrating Legacy Software to Microservices Architecture,'' 2019 Open Conference of Electrical, Electronic and Information Sciences (eStream), Vilnius, Lithuania, 2019, pp. 1-5, doi: 10.1109/eStream.2019.8732170. \\
			\hline
			LR5 & J. Fritzsch, J. Bogner, S. Wagner and A. Zimmermann, ``Microservices Migration in Industry: Intentions, Strategies, and Challenges,'' 2019 IEEE International Conference on Software Maintenance and Evolution (ICSME), Cleveland, OH, USA, 2019, pp. 481-490, doi: 10.1109/ICSME.2019.00081. \\
			\hline
			LR6 & R. Capuano and H. Muccini, ``A Systematic Literature Review on Migration to Microservices: a Quality Attributes perspective,'' 2022 IEEE 19th International Conference on Software Architecture Companion (ICSA-C), Honolulu, HI, USA, 2022, pp. 120-123, doi: 10.1109/ICSA-C54293.2022.00030. \\
			\hline
			LR7 & F. Ponce, G. Márquez and H. Astudillo, ``Migrating from monolithic architecture to microservices: A Rapid Review,'' 2019 38th International Conference of the Chilean Computer Science Society (SCCC), Concepcion, Chile, 2019, pp. 1-7, doi: 10.1109/SCCC49216.2019.8966423. \\
			\hline
			LR8 & Fritzsch, J., Bogner, J., Zimmermann, A., Wagner, S. (2019). From Monolith to Microservices: A Classification of Refactoring Approaches. In: Bruel, JM., Mazzara, M., Meyer, B. (eds) Software Engineering Aspects of Continuous Development and New Paradigms of Software Production and Deployment. DEVOPS 2018. Lecture Notes in Computer Science(), vol 11350. Springer, Cham. https://doi.org/10.1007/978-3-030-06019-0\_10 \\
			\hline
			LR9 & Taibi, D.; Lenarduzzi, V. and Pahl, C. (2018). Architectural Patterns for Microservices: A Systematic Mapping Study. In Proceedings of the 8th International Conference on Cloud Computing and Services Science - CLOSER; ISBN 978-989-758-295-0; ISSN 2184-5042, SciTePress, pages 221-232. DOI: 10.5220/0006798302210232 \\
			\hline
			LR10 & L. Carvalho, A. Garcia, W. K. G. Assunção, R. de Mello and M. Julia de Lima, ``Analysis of the Criteria Adopted in Industry to Extract Microservices,'' 2019 IEEE/ACM Joint 7th International Workshop on Conducting Empirical Studies in Industry (CESI) and 6th International Workshop on Software Engineering Research and Industrial Practice (SER\&IP), Montreal, QC, Canada, 2019, pp. 22-29, doi: 10.1109/CESSER-IP.2019.00012. \\
			\hline
			LR11 & Daniele Wolfart, Wesley K. G. Assunção, Ivonei F. da Silva, Diogo C. P. Domingos, Ederson Schmeing, Guilherme L. Donin Villaca, and Diogo do N. Paza. 2021. Modernizing Legacy Systems with Microservices: A Roadmap. In Evaluation and Assessment in Software Engineering (EASE 2021). Association for Computing Machinery, New York, NY, USA, 149–159. https://doi.org/10.1145/3463274.3463334. \\
			\hline
			LR12 & Abdellatif, M., Shatnawi, A., Mili, H., Moha, N., Boussaidi, G. el, Hecht, G., Privat, J., \& Guéhéneuc, Y. G. (2021). A taxonomy of service identification approaches for legacy software systems modernization. Journal of Systems and Software, 173, 110868. https://doi.org/10.1016/J.JSS.2020.110868 \\
			\hline
		\end{longtable}
}
	
\section{Queries for Paper Selection}
\label{appendix:queries}
	
	\begin{table}[H]
		\noindent\adjustbox{max width=\textwidth}{%
        \begin{tabular}{|l|l|l|}
        \hline
        \multicolumn{1}{|c|}{\textbf{Research paper library}} & \multicolumn{1}{c|}{\textbf{Search query}} & \multicolumn{1}{c|}{\textbf{Result count}} \\ \hline
        Web of Science & \begin{tabular}[c]{@{}l@{}}(TS=(microservice* AND (reengineer* OR re-engineer* OR redesign* OR re-design* \\ OR discover* OR identify*))) AND  (WC=(Computer Science)) AND (DT=(Article OR \\ Book Chapter OR Proceedings Paper)) AND  (LA=(English))\end{tabular} & 336 \\ \hline
        Scopus & \begin{tabular}[c]{@{}l@{}}( TITLE-ABS-KEY ( microservice*  AND  ( reengineer*  OR  re-engineer*  OR  redesign*  \\ OR  re-design*  OR  discover*  OR  identify* ) ) )  AND  ( LIMIT-TO ( DOCTYPE ,  "cp" ) \\  OR  LIMIT-TO ( DOCTYPE ,  "ar" )  OR  LIMIT-TO ( DOCTYPE ,  "ch" ) )  AND  \\ ( LIMIT-TO ( SUBJAREA ,  "COMP" ) )  AND  ( LIMIT-TO ( LANGUAGE ,  "English" ) )\end{tabular} & 522 \\ \hline
        ScienceDirect & \begin{tabular}[c]{@{}l@{}}Title, abstract or author-specified keywords: "(microservice AND (reengineer OR re-engineer \\ OR redesign OR re-design OR discover OR identify))"; Filter by Article Type Research Articles \\ AND Subject Areas Computer Science\end{tabular} & 41 \\ \hline
        \begin{tabular}[c]{@{}l@{}}ACM DL \\ Content type - Survey\end{tabular} & \begin{tabular}[c]{@{}l@{}}{[}Full Text: microservice*{]} AND {[}{[}Full Text: reengineer*{]} OR {[}Full Text: re-engineer*{]} OR \\ {[}Full Text: redesign*{]} OR {[}Full Text: re-design*{]} OR {[}Full Text: discover*{]} OR \\ {[}Full Text: identify*{]}{]}\end{tabular} & 1332 \\ \hline
        \begin{tabular}[c]{@{}l@{}}ACM DL\\ Content type - \\ Research Article\end{tabular} & \begin{tabular}[c]{@{}l@{}}{[}Full Text: microservice*{]} AND {[}{[}Full Text: reengineer*{]} OR {[}Full Text: re-engineer*{]} OR \\ {[}Full Text: redesign*{]} OR {[}Full Text: re-design*{]} OR {[}Full Text: discover*{]} OR \\ {[}Full Text: identify*{]}{]}\end{tabular} & 1290 \\ \hline
        IEEE Xplore & \begin{tabular}[c]{@{}l@{}}(("All Metadata":microservice*) AND ("All Metadata":redesign* OR "All Metadata":re-design* \\ OR "All Metadata":reengineer* OR "All Metadata":re-engineer* OR "All Metadata":identify* \\ OR "All Metadata":discover*)) Filters Applied: Conferences Journals\end{tabular} & 257 \\ \hline
        \end{tabular}}
	\end{table}

\section{Study List}
\label{appendix:studyList}
{
\scriptsize		
  \begin{longtable}{|l|p{0.93\textwidth}|}

    		\hline
			\textbf{ID} & \textbf{Full reference} \\
			\hline
				1 & Gysel, M., Kölbener, L., Giersche, W., \& Zimmermann, O. (2016). Service Cutter: A Systematic Approach to Service Decomposition. European Conference on Service-Oriented and Cloud Computing. \\ \hline
				2 & Taibi, Davide \& Systä, Kari. (2019). From Monolithic Systems to Microservices: A Decomposition Framework based on Process Mining. 10.5220/0007755901530164. \\ \hline
				3 & Florian Auer, Valentina Lenarduzzi, Michael Felderer, Davide Taibi, From monolithic systems to Microservices: An assessment framework, Information and Software Technology, Volume 137, 2021, 106600, ISSN 0950-5849, https://doi.org/10.1016/j.infsof.2021.106600. \\ \hline
				4 & Hamdy Michael Ayas, Philipp Leitner, and Regina Hebig. 2021. The Migration Journey Towards Microservices. In Product-Focused Software Process Improvement: 22nd International Conference, PROFES 2021, Turin, Italy, November 26, 2021, Proceedings. Springer-Verlag, Berlin, Heidelberg, 20–35. https://doi.org/10.1007/978-3-030-91452-3\_2 \\ \hline
				5 & Bajaj, Deepali \& Bharti, Urmil \& Goel, Anita \& Gupta, S.. (2020). Partial Migration for Re-architecting a Cloud Native Monolithic Application into Microservices and FaaS. 10.1007/978-981-15-9671-1\_9. \\ \hline
				6 & Abdullah, Muhammad \& Iqbal, Waheed \& Erradi, Abdelkarim. (2019). Unsupervised Learning Approach for Web Application  Auto-Decomposition into Microservices. Journal of Systems and Software. 151. 10.1016/j.jss.2019.02.031. \\ \hline
				7 & Bajaj, Deepali \& Goel, Anita \& Gupta, S.. (2022). GreenMicro: Identifying Microservices From Use Cases in Greenfield Development. IEEE Access. 10. 1-1. 10.1109/ACCESS.2022.3182495. \\ \hline
				8 & Barbosa, Marx \& Maia, Paulo. (2020). Towards Identifying Microservice Candidates from Business Rules Implemented in Stored Procedures. 10.1109/ICSA-C50368.2020.00015. \\ \hline
				9 & Belafia, R., Jeanjean, P., Barais, O., Guernic, G.L., \& Combemale, B. (2021). From Monolithic to  Microservice Architecture: The Case of Extensible and Domain-Specific IDEs. 2021 ACM/IEEE International  Conference on Model Driven Engineering Languages and Systems Companion (MODELS-C), 454-463. \\ \hline
				10 & Bozan, Karoly \& Lyytinen, Kalle \& Rose, Gregory. (2020). How to transition incrementally to microservice architecture. Communications of the ACM. 64. 79-85. 10.1145/3378064. \\ \hline
				11 & Miguel Brito, Jácome Cunha, and João Saraiva. 2021. Identification of microservices from monolithic applications through topic modelling. In Proceedings of the 36th Annual ACM Symposium on Applied Computing (SAC '21). Association for Computing Machinery, New York, NY, USA, 1409–1418. https://doi.org/10.1145/3412841.3442016 \\ \hline
				12 & Bushong, Vincent \& Das, Dipta \& Maruf, Abdullah \& Černý, Tom. (2021). Using Static Analysis to Address Microservice Architecture  Reconstruction. 1199-1201. 10.1109/ASE51524.2021.9678749. \\ \hline
				13 & Lingli Cao and Cheng Zhang. 2022. Implementation of Domain-oriented Microservices Decomposition based on Node-attributed Network. In Proceedings of the 2022 11th International Conference on Software and Computer Applications (ICSCA '22). Association for Computing Machinery, New York, NY, USA, 136–142. https://doi.org/10.1145/3524304.3524325 \\ \hline
				14 & Andrés Carrasco, Brent van Bladel, and Serge Demeyer. 2018. Migrating towards microservices: migration and architecture smells. In Proceedings of the 2nd International Workshop on Refactoring (IWoR 2018). Association for Computing Machinery, New York, NY, USA, 1–6. https://doi.org/10.1145/3242163.3242164 \\ \hline
				15 & L. Carvalho, A. Garcia, W. K. G. Assunção, R. de Mello and M. Julia de Lima, "Analysis of the Criteria Adopted in Industry to Extract Microservices," 2019 IEEE/ACM Joint 7th International Workshop on Conducting Empirical Studies in Industry (CESI) and 6th International Workshop on Software Engineering Research and Industrial Practice (SER\&IP), Montreal, QC, Canada, 2019, pp. 22-29, doi: 10.1109/CESSER-IP.2019.00012. \\ \hline
				16 & Tomas Cerny, Filip Sedlisky, and Michael J. Donahoo. 2018. On isolation-driven automated module decomposition. In Proceedings of the 2018 Conference on Research in Adaptive and Convergent Systems (RACS '18). Association for Computing Machinery, New York, NY, USA, 302–307. https://doi.org/10.1145/3264746.3264756 \\ \hline
				17 & Nacha Chondamrongkul, Jing Sun, and Ian Warren. 2021. Software Architectural Migration: An Automated Planning Approach. ACM Trans. Softw. Eng. Methodol. 30, 4, Article 50 (October 2021), 35 pages. https://doi.org/10.1145/3461011 \\ \hline
				18 & Christoforou, Andreas \& Garriga, Martin \& Andreou, Andreas \& Baresi, Luciano. (2017). Supporting the Decision of  Migrating to Microservices Through Multi-layer Fuzzy Cognitive Maps. 471-480. 10.1007/978-3-319-69035-3\_34. \\ \hline
				19 & Carlos Eduardo da Silva, Yan de Lima Justino, and Eiji Adachi. 2022. SPReaD: service-oriented process for reengineering and DevOps: Developing microservices for a Brazilian state department of taxation. Serv. Oriented Comput. Appl. 16, 1 (Mar 2022), 1–16. https://doi.org/10.1007/s11761-021-00329-x \\ \hline
				20 & Hugo S. da Silva, Glauco Carneiro, and Miguel Monteiro. 2023. Towards a Roadmap for the Migration of Legacy Software Systems to a Microservice based Architecture. In Proceedings of the 9th International Conference on Cloud Computing and Services Science (CLOSER 2019). SCITEPRESS - Science and Technology Publications, Lda, Setubal, PRT, 37–47. https://doi.org/10.5220/0007618400370047 \\ \hline
				21 & M. Daoud, A. El Mezouari, N. Faci, D. Benslimane, Z. Maamar and A. El Fazziki, "Towards an Automatic Identification of Microservices from Business Processes," 2020 IEEE 29th International Conference on Enabling Technologies: Infrastructure for Collaborative Enterprises (WETICE), Bayonne, France, 2020, pp. 42-47, doi: 10.1109/WETICE49692.2020.00017. \\ \hline
				22 & Dattatreya, V., Chalapati Rao, K.V., Raghava, M. (2021). Design Patterns and Microservices for Reengineering of Legacy Web Applications. In: Satapathy, S.C., Bhateja, V., Favorskaya, M.N., Adilakshmi, T. (eds) Smart Computing Techniques and Applications. Smart Innovation, Systems and Technologies, vol 224. Springer, Singapore. https://doi.org/10.1007/978-981-16-1502-3\_38 \\ \hline
				23 & de Almeida, M.G., Canedo, E.D. (2022). The Adoption of Microservices Architecture as a Natural Consequence of Legacy System Migration at Police Intelligence Department. In: Gervasi, O., Murgante, B., Hendrix, E.M.T., Taniar, D., Apduhan, B.O. (eds) Computational Science and Its Applications – ICCSA 2022. ICCSA 2022. Lecture Notes in Computer Science, vol 13375. Springer, Cham. https://doi.org/10.1007/978-3-031-10522-7\_25 \\ \hline
				24 & De Alwis, A.A.C., Barros, A., Fidge, C., Polyvyanyy, A. (2018). Discovering Microservices in Enterprise Systems Using a Business Object Containment Heuristic. In: Panetto, H., Debruyne, C., Proper, H., Ardagna, C., Roman, D., Meersman, R. (eds) On the Move to Meaningful Internet Systems. OTM 2018 Conferences. OTM 2018. Lecture Notes in Computer Science(), vol 11230. Springer, Cham. https://doi.org/10.1007/978-3-030-02671-4\_4 \\ \hline
				25 & De Alwis, A.A.C., Barros, A., Polyvyanyy, A., Fidge, C. (2018). Function-Splitting Heuristics for Discovery of Microservices in Enterprise Systems. In: Pahl, C., Vukovic, M., Yin, J., Yu, Q. (eds) Service-Oriented Computing. ICSOC 2018. Lecture Notes in Computer Science(), vol 11236. Springer, Cham. https://doi.org/10.1007/978-3-030-03596-9\_3 \\ \hline
				26 & De Alwis, A.A.C., Barros, A., Fidge, C., Polyvyanyy, A. (2019). Availability and Scalability Optimized Microservice Discovery from Enterprise Systems. In: Panetto, H., Debruyne, C., Hepp, M., Lewis, D., Ardagna, C., Meersman, R. (eds) On the Move to Meaningful Internet Systems: OTM 2019 Conferences. OTM 2019. Lecture Notes in Computer Science(), vol 11877. Springer, Cham. https://doi.org/10.1007/978-3-030-33246-4\_31 \\ \hline
				27 & De Alwis, A.A.C., Barros, A., Fidge, C., Polyvyanyy, A. (2020). Remodularization Analysis for Microservice Discovery Using Syntactic and Semantic Clustering. In: Dustdar, S., Yu, E., Salinesi, C., Rieu, D., Pant, V. (eds) Advanced Information Systems Engineering. CAiSE 2020.Lecture Notes in Computer Science(), vol 12127. Springer, Cham. https://doi.org/10.1007/978-3-030-49435-3\_1 \\ \hline
				28 & De Alwis, A.A.C., Barros, A., Fidge, C., Polyvyanyy, A. (2021). Microservice Remodularisation of Monolithic Enterprise Systems for Embedding in Industrial IoT Networks. In: La Rosa, M., Sadiq, S., Teniente, E. (eds) Advanced Information Systems Engineering. CAiSE 2021. Lecture Notes in Computer Science(), vol 12751. Springer, Cham. https://doi.org/10.1007/978-3-030-79382-1\_26 \\ \hline
				29 & Desai, U., Bandyopadhyay, S., \& Tamilselvam, S.G. (2021). Graph Neural Network to Dilute Outliers for Refactoring Monolith Application. AAAI Conference on Artificial Intelligence. \\ \hline
				30 & E. Djogic, S. Ribic and D. Donko, "Monolithic to microservices redesign of event driven integration platform," 2018 41st International Convention on Information and Communication Technology, Electronics and Microelectronics (MIPRO), Opatija, Croatia, 2018, pp. 1411-1414, doi: 10.23919/MIPRO.2018.8400254. \\ \hline
				31 & D. Escobar et al., "Towards the understanding and evolution of monolithic applications as microservices," 2016 XLII Latin American Computing Conference (CLEI), Valparaiso, Chile, 2016, pp. 1-11, doi: 10.1109/CLEI.2016.7833410. \\ \hline
				32 & Sinan Eski and Feza Buzluca. 2018. An automatic extraction approach: transition to microservices architecture from monolithic application. In Proceedings of the 19th International Conference on Agile Software Development: Companion (XP '18). Association for Computing Machinery, New York, NY, USA, Article 25, 1–6. https://doi.org/10.1145/3234152.3234195 \\ \hline
				33 & Francisco Freitas, André Ferreira, and Jácome Cunha. 2021. Refactoring Java Monoliths into Executable Microservice-Based Applications. In Proceedings of the 25th Brazilian Symposium on Programming Languages (SBLP '21). Association for Computing Machinery, New York, NY, USA, 100–107. https://doi.org/10.1145/3475061.3475086 \\ \hline
				34 & A. Furda, C. Fidge, O. Zimmermann, W. Kelly and A. Barros, "Migrating Enterprise Legacy Source Code to Microservices: On Multitenancy, Statefulness, and Data Consistency," in IEEE Software, vol. 35, no. 3, pp. 63-72, May/June 2018, doi: 10.1109/MS.2017.440134612. \\ \hline
				35 & Gutiérrez-Fernández, A.M., Resinas, M., \& Ruiz-Cortés, A. (2016). Redefining a Process Engine as a Microservice Platform. Business Process Management Workshops. \\ \hline
				36 & A. O. R. Ishida, K. Kontogiannis and C. Brealey, "Extracting Micro Service Dependencies Using Log Analysis," 2022 IEEE 29th Annual Software Technology Conference (STC), Gaithersburg, MD, USA, 2022, pp. 82-92, doi: 10.1109/STC55697.2022.00020. \\ \hline
				37 & Md Rofiqul Islam and Tomas Cerny. 2022. Business process extraction using static analysis. In Proceedings of the 36th IEEE/ACM International Conference on Automated Software Engineering (ASE '21). IEEE Press, 1202–1204. https://doi.org/10.1109/ASE51524.2021.9678588 \\ \hline
				38 & A. Janes and B. Russo, "Automatic Performance Monitoring and Regression Testing During the Transition from Monolith to Microservices," 2019 IEEE International Symposium on Software Reliability Engineering Workshops (ISSREW), Berlin, Germany, 2019, pp. 163-168, doi: 10.1109/ISSREW.2019.00067. \\ \hline
				39 & W. Jin, T. Liu, Y. Cai, R. Kazman, R. Mo and Q. Zheng, "Service Candidate Identification from Monolithic Systems Based on Execution Traces," in IEEE Transactions on Software Engineering, vol. 47, no. 5, pp. 987-1007, 1 May 2021, doi: 10.1109/TSE.2019.2910531. \\ \hline
				40 & M. I. Joselyne, G. Bajpai and F. Nzanywayingoma, "A Systematic Framework of Application Modernization to Microservice based Architecture," 2021 International Conference on Engineering and Emerging Technologies (ICEET), Istanbul, Turkey, 2021, pp. 1-6, doi: 10.1109/ICEET53442.2021.9659783. \\ \hline
				41 & I. J. Munezero, D. -T. Mukasa, B. Kanagwa and J. Balikuddembe, "Partitioning Microservices: A Domain Engineering Approach," 2018 IEEE/ACM Symposium on Software Engineering in Africa (SEiA), Gothenburg, Sweden, 2018, pp. 43-49. \\ \hline
				42 & Anup K. Kalia, Jin Xiao, Rahul Krishna, Saurabh Sinha, Maja Vukovic, and Debasish Banerjee. 2021. Mono2Micro: a practical and effective tool for decomposing monolithic Java applications to microservices. In Proceedings of the 29th ACM Joint Meeting on European Software Engineering Conference and Symposium on the Foundations of Software Engineering (ESEC/FSE 2021). Association for Computing Machinery, New York, NY, USA, 1214–1224. https://doi.org/10.1145/3468264.3473915 \\ \hline
				43 & Anup K. Kalia, Jin Xiao, Chen Lin, Saurabh Sinha, John Rofrano, Maja Vukovic, and Debasish Banerjee. 2020. Mono2Micro: an AI-based toolchain for evolving monolithic enterprise applications to a microservice architecture. In Proceedings of the 28th ACM Joint Meeting on European Software Engineering Conference and Symposium on the Foundations of Software Engineering (ESEC/FSE 2020). Association for Computing Machinery, New York, NY, USA, 1606–1610. https://doi.org/10.1145/3368089.3417933 \\ \hline
				44 & M. Kamimura, K. Yano, T. Hatano and A. Matsuo, "Extracting Candidates of Microservices from Monolithic Application Code," 2018 25th Asia-Pacific Software Engineering Conference (APSEC), Nara, Japan, 2018, pp. 571-580, doi: 10.1109/APSEC.2018.00072. \\ \hline
				45 & Khoshnevis, S. A search-based identification of variable microservices for enterprise SaaS. Front. Comput. Sci. 17, 173208 (2023). https://doi.org/10.1007/s11704-022-1390-4 \\ \hline
				46 & A. Krause, C. Zirkelbach, W. Hasselbring, S. Lenga and D. Kroger, "Microservice Decomposition via Static and Dynamic Analysis of the Monolith," in 2020 IEEE International Conference on Software Architecture Companion (ICSA-C), Salvador, Brazil, 2020 pp. 9-16.doi: 10.1109/ICSA-C50368.2020.00011 \\ \hline
				47 & Lapuz, N., Clarke, P., Abgaz, Y. (2021). Digital Transformation and the Role of Dynamic Tooling in Extracting Microservices from Existing Software Systems. In: Yilmaz, M., Clarke, P., Messnarz, R., Reiner, M. (eds) Systems, Software and Services Process Improvement. EuroSPI 2021. Communications in Computer and Information Science, vol 1442. Springer, Cham. https://doi.org/10.1007/978-3-030-85521-5\_20 \\ \hline
				48 & Shanshan Li, He Zhang, Zijia Jia, Zheng Li, Cheng Zhang, Jiaqi Li, Qiuya Gao, Jidong Ge, Zhihao Shan,A dataflow-driven approach to identifying microservices from monolithic applications, Journal of Systems and Software, Volume 157, 2019, 110380, ISSN 0164-1212, https://doi.org/10.1016/j.jss.2019.07.008. \\ \hline
				49 & Zhiding Li, Chenqi Shang, Jianjie Wu, Yuan Li, Microservice extraction based on knowledge graph from monolithic applications, Information and Software Technology, Volume 150, 2022, 106992, ISSN 0950-5849, https://doi.org/10.1016/j.infsof.2022.106992. \\ \hline
				50 & B. Liu, J. Lu, F. Zhang, W. Zhang and M. Wang, "Method of Microservices Division for Complex Business Management System Based on Dual Clustering," 2020 5th International Conference on Mechanical, Control and Computer Engineering (ICMCCE), Harbin, China, 2020, pp. 2259-2268, doi: 10.1109/ICMCCE51767.2020.00490. \\ \hline
				51 & Löhnertz, J., \& Oprescu, A. (2020). Steinmetz: Toward Automatic Decomposition of Monolithic Software Into Microservices. Seminar on Advanced Techniques and Tools for Software Evolution. \\ \hline
				52 & Matias, T., Correia, F.F., Fritzsch, J., Bogner, J., Ferreira, H.S., Restivo, A. (2020). Determining Microservice Boundaries: A Case Study Using Static and Dynamic Software Analysis. In: Jansen, A., Malavolta, I., Muccini, H., Ozkaya, I., Zimmermann, O. (eds) Software Architecture. ECSA 2020. Lecture Notes in Computer Science(), vol 12292. Springer, Cham. https://doi.org/10.1007/978-3-030-58923-3\_21 \\ \hline
				53 & Vikram Nitin, Shubhi Asthana, Baishakhi Ray, and Rahul Krishna. 2023. CARGO: AI-Guided Dependency Analysis for Migrating Monolithic Applications to Microservices Architecture. In Proceedings of the 37th IEEE/ACM International Conference on Automated Software Engineering (ASE '22). Association for Computing Machinery, New York, NY, USA, Article 20, 1–12. https://doi.org/10.1145/3551349.3556960 \\ \hline
				54 & park, J., Moon, M. and Keunhyuk, K. 2019. Approach to Identify Microservices based on Analysis Class Model. International Journal of Advanced Science and Technology. 28, 4 (Sep. 2019), 08 - 14. \\ \hline
				55 & Pigazzini, I., Fontana, F.A., \& Maggioni, A. (2019). Tool Support for the Migration to Microservice Architecture: An Industrial Case Study. European Conference on Software Architecture. \\ \hline
				56 & T. Prasandy, Titan, D. F. Murad and T. Darwis, "Migrating Application from Monolith to Microservices," 2020 International Conference on Information Management and Technology (ICIMTech), Bandung, Indonesia, 2020, pp. 726-731, doi: 10.1109/ICIMTech50083.2020.9211252. \\ \hline
				57 & Zhongshan Ren, Wei Wang, Guoquan Wu, Chushu Gao, Wei Chen, Jun Wei, and Tao Huang. 2018. Migrating Web Applications from Monolithic Structure to Microservices Architecture. In Proceedings of the 10th Asia-Pacific Symposium on Internetware (Internetware '18). Association for Computing Machinery, New York, NY, USA, Article 7, 1–10. https://doi.org/10.1145/3275219.3275230 \\ \hline
				58 & Y. Romani, O. Tibermacine and C. Tibermacine, "Towards Migrating Legacy Software Systems to Microservice-based Architectures: a Data-Centric Process for Microservice Identification," 2022 IEEE 19th International Conference on Software Architecture Companion (ICSA-C), Honolulu, HI, USA, 2022, pp. 15-19, doi: 10.1109/ICSA-C54293.2022.00010. \\ \hline
				59 & Saidi, M., Tissaoui, A., Benslimane, D., Faiz, S. (2022). Automatic Microservices Identification Across Structural Dependency. In: Abraham, A., et al. Hybrid Intelligent Systems. HIS 2021. Lecture Notes in Networks and Systems, vol 420. Springer, Cham. https://doi.org/10.1007/978-3-030-96305-7\_36 \\ \hline
				60 & N. Santos and A. Rito Silva, "A Complexity Metric for Microservices Architecture Migration," 2020 IEEE International Conference on Software Architecture (ICSA), Salvador, Brazil, 2020, pp. 169-178, doi: 10.1109/ICSA47634.2020.00024. \\ \hline
				61 & Santos, Samuel \& Silva, António. (2022). Microservices Identification in Monolith Systems: Functionality Redesign Complexity and Evaluation of Similarity Measures. Journal of Web Engineering. 10.13052/jwe1540-9589.2158. \\ \hline
				62 & Sarita and S. Sebastian, "Transform Monolith into Microservices using Docker," 2017 International Conference on Computing, Communication, Control and Automation (ICCUBEA), Pune, India, 2017, pp. 1-5, doi: 10.1109/ICCUBEA.2017.8463820. \\ \hline
				63 & Casper Schröder, Adriaan van der Feltz, Annibale Panichella, and Maurício Aniche. 2021. Search-based software re-modularization: a case study at Adyen. In Proceedings of the 43rd International Conference on Software Engineering: Software Engineering in Practice (ICSE-SEIP '21). IEEE Press, 81–90. https://doi.org/10.1109/ICSE-SEIP52600.2021.00017 \\ \hline
				64 & Christoph Schröer, Towards Microservice Identification Approaches for Architecting Data Science Workflows, Procedia Computer Science, Volume 181, 2021, Pages 519-525, \\ \hline
				65 & C. Schroer, S. Wittfoth and J. M. Gomez, "A Process Model for Microservices Design and Identification," 2021 IEEE 18th International Conference on Software Architecture Companion (ICSA-C), Stuttgart, Germany, 2021, pp. 1-8, doi: 10.1109/ICSA-C52384.2021.00013. \\ \hline
				66 & Khaled Sellami, Mohamed Aymen Saied, and Ali Ouni. 2022. A Hierarchical DBSCAN Method for Extracting Microservices from Monolithic Applications. In Proceedings of the 26th International Conference on Evaluation and Assessment in Software Engineering (EASE '22). Association for Computing Machinery, New York, NY, USA, 201–210. https://doi.org/10.1145/3530019.3530040 \\ \hline
				67 & Selmadji, A., Seriai, AD., Bouziane, H.L., Dony, C., Mahamane, R.O. (2018). Re-architecting OO Software into Microservices. In: Kritikos, K., Plebani, P., de Paoli, F. (eds) Service-Oriented and Cloud Computing. ESOCC 2018. Lecture Notes in Computer Science(), vol 11116. Springer, Cham. https://doi.org/10.1007/978-3-319-99819-0\_5 \\ \hline
				68 & A. L. Shastry, D. S. Nair, B. Prathima, C. P. Ramya and P. Hallymysore, "Approaches for migrating non cloud-native applications to the cloud," 2022 IEEE 12th Annual Computing and Communication Workshop and Conference (CCWC), Las Vegas, NV, USA, 2022, pp. 0632-0638, doi: 10.1109/CCWC54503.2022.9720856. \\ \hline
				69 & T. D. Stojanovic, S. D. Lazarevic, M. Milic and I. Antovic, "Identifying microservices using structured system analysis," 2020 24th International Conference on Information Technology (IT), Zabljak, Montenegro, 2020, pp. 1-4, doi: 10.1109/IT48810.2020.9070652. \\ \hline
				70 & Sun, Xiaoxiao \& Boranbaev, Salamat \& Han, Shicong \& Wang, Huanqiang \& Yu, Dongjin. (2022). Expert system for automatic microservices identification using API similarity graph. Expert Systems. 10.1111/exsy.13158. \\ \hline
				71 & D. Taibi, V. Lenarduzzi and C. Pahl, "Processes, Motivations, and Issues for Migrating to Microservices Architectures: An Empirical Investigation," in IEEE Cloud Computing, vol. 4, no. 5, pp. 22-32, September/October 2017, doi: 10.1109/MCC.2017.4250931. \\ \hline
				72 & Taibi, D., Systä, K. (2020). A Decomposition and Metric-Based Evaluation Framework for Microservices. In: Ferguson, D., Méndez Muñoz, V., Pahl, C., Helfert, M. (eds) Cloud Computing and Services Science. CLOSER 2019. Communications in Computer and Information Science, vol 1218. Springer, Cham. https://doi.org/10.1007/978-3-030-49432-2\_7 \\ \hline
				73 & M. Tusjunt and W. Vatanawood, "Refactoring Orchestrated Web Services into Microservices Using Decomposition Pattern," 2018 IEEE 4th International Conference on Computer and Communications (ICCC), Chengdu, China, 2018, pp. 609-613, doi: 10.1109/CompComm.2018.8781036. \\ \hline
				74 & Tyszberowicz, S., Heinrich, R., Liu, B., Liu, Z. (2018). Identifying Microservices Using Functional Decomposition. In: Feng, X., Müller-Olm, M., Yang, Z. (eds) Dependable Software Engineering. Theories, Tools, and Applications. SETTA 2018. Lecture Notes in Computer Science(), vol 10998. Springer, Cham. https://doi.org/10.1007/978-3-319-99933-3\_4 \\ \hline
				75 & Vera-Rivera, F.H., Puerto-Cuadros, E.G., Astudillo, H., Gaona-Cuevas, C.M. (2020). Microservices Backlog - A Model of Granularity Specification and Microservice Identification. In: Wang, Q., Xia, Y., Seshadri, S., Zhang, LJ. (eds) Services Computing – SCC 2020. SCC 2020. Lecture Notes in Computer Science(), vol 12409. Springer, Cham. https://doi.org/10.1007/978-3-030-59592-0\_6 \\ \hline
				76 & E. Volynsky, M. Mehmed and S. Krusche, "Architect: A Framework for the Migration to Microservices," 2022 International Conference on Computing, Electronics \& Communications Engineering (iCCECE), Southend, United Kingdom, 2022, pp. 71-76, doi: 10.1109/iCCECE55162.2022.9875096. \\ \hline
				77 & Yuyang Wei, Yijun Yu, Minxue Pan, and Tian Zhang. 2021. A Feature Table approach to decomposing monolithic applications into microservices. In Proceedings of the 12th Asia-Pacific Symposium on Internetware (Internetware '20). Association for Computing Machinery, New York, NY, USA, 21–30. https://doi.org/10.1145/3457913.3457939 \\ \hline
				78 & R. Yedida, R. Krishna, A. Kalia, T. Menzies, J. Xiao and M. Vukovic, "Lessons learned from hyper-parameter tuning for microservice candidate identification," 2021 36th IEEE/ACM International Conference on Automated Software Engineering (ASE), Melbourne, Australia, 2021, pp. 1141-1145, doi: 10.1109/ASE51524.2021.9678704. \\ \hline
				79 & P. Zaragoza, A. -D. Seriai, A. Seriai, A. Shatnawi and M. Derras, "Leveraging the Layered Architecture for Microservice Recovery," 2022 IEEE 19th International Conference on Software Architecture (ICSA), Honolulu, HI, USA, 2022, pp. 135-145, doi: 10.1109/ICSA53651.2022.00021. \\ \hline
				80 & Al-Debagy, Omar and Peter Martinek. “Dependencies-based microservices decomposition method.” International Journal of Computers and Applications 44 (2021): 814 - 821. \\ \hline
				81 & Almeida, J.F., Silva, A.R. (2020). Monolith Migration Complexity Tuning Through the Application of Microservices Patterns. In: Jansen, A., Malavolta, I., Muccini, H., Ozkaya, I., Zimmermann, O. (eds) Software Architecture. ECSA 2020. Lecture Notes in Computer Science(), vol 12292. Springer, Cham. https://doi.org/10.1007/978-3-030-58923-3\_3 \\ \hline
				82 & Assunção, W.K.G., Colanzi, T.E., Carvalho, L. et al. Analysis of a many-objective optimization approach for identifying microservices from legacy systems. Empir Software Eng 27, 51 (2022). https://doi.org/10.1007/s10664-021-10049-7 \\ \hline
				83 & W. K. G. Assunção et al., "A Multi-Criteria Strategy for Redesigning Legacy Features as Microservices: An Industrial Case Study," 2021 IEEE International Conference on Software Analysis, Evolution and Reengineering (SANER), Honolulu, HI, USA, 2021, pp. 377-387, doi: 10.1109/SANER50967.2021.00042. \\ \hline
				84 & Wesley K. G. Assunção, Jacob Krüger, and Willian D. F. Mendonça. 2020. Variability management meets microservices: six challenges of re-engineering microservice-based webshops. In Proceedings of the 24th ACM Conference on Systems and Software Product Line: Volume A - Volume A (SPLC '20). Association for Computing Machinery, New York, NY, USA, Article 22, 1–6. https://doi.org/10.1145/3382025.3414942 \\ \hline
				85 & Baresi, L., Garriga, M., De Renzis, A. (2017). Microservices Identification Through Interface Analysis. In: De Paoli, F., Schulte, S., Broch Johnsen, E. (eds) Service-Oriented and Cloud Computing. ESOCC 2017. Lecture Notes in Computer Science(), vol 10465. Springer, Cham. https://doi.org/10.1007/978-3-319-67262-5\_2 \\ \hline
				86 & G. Mazlami, J. Cito and P. Leitner, "Extraction of Microservices from Monolithic Software Architectures," in 2017 IEEE International Conference on Web Services (ICWS), Honolulu, HI, 2017 pp. 524-531. doi: 10.1109/ICWS.2017.61 \\ \hline
				87 & Al-Debagy, Omar \& Martinek, Peter. (2019). A New Decomposition Method for Designing Microservices. Periodica Polytechnica Electrical Engineering and Computer Science. 63. 10.3311/PPee.13925. \\ \hline
				88 & W. Jin, T. Liu, Q. Zheng, D. Cui and Y. Cai, "Functionality-Oriented Microservice Extraction Based on Execution Trace Clustering," 2018 IEEE International Conference on Web Services (ICWS), San Francisco, CA, USA, 2018, pp. 211-218, doi: 10.1109/ICWS.2018.00034. \\ \hline
				89 & Trabelsi, Imen \& Abdellatif, Manel \& Abubaker, Abdalgader \& Moha, Naouel \& Mosser, Sébastien \& Ebrahimi‐Kahou, Samira \& Guéhéneuc, Yann-Gaël. (2022). From legacy to microservices: A type‐based approach for microservices identification using machine learning and semantic analysis. Journal of Software: Evolution and Process. 10.1002/smr.2503. \\ \hline
				90 & Al-Debagy, Omar. “A Microservice Decomposition Method Through Using Distributed Representation of Source Code.” Scalable Comput. Pract. Exp. 22 (2021): 39-52. \\ \hline
				91 & Levcovitz, A., Terra, R., \& Valente, M. T. (2016). Towards a Technique for Extracting Microservices from Monolithic Enterprise Systems. ArXiv. /abs/1605.03175 \\ \hline
				92 & M. J. Amiri, "Object-Aware Identification of Microservices," 2018 IEEE International Conference on Services Computing (SCC), San Francisco, CA, USA, 2018, pp. 253-256, doi: 10.1109/SCC.2018.00042. \\ \hline
			\end{longtable}
}

\end{document}